\documentclass[twocolumn,showpacs,preprintnumbers,amsmath,amssymb,epsfig,widetext]{revtex4}
\usepackage{graphicx}
\usepackage{dcolumn}
\usepackage{bm}
\usepackage{epsfig}
\usepackage{color}

\usepackage{hyperref}

\hypersetup{
    colorlinks=true,
    linkcolor=red,
    citecolor=blue,
}

\def\la{\mathrel{\mathpalette\fun <}}

\def\fun#1#2{\lower3.6pt\vbox{\baselineskip0pt\lineskip.9pt
  \ialign{$\mathsurround=0pt#1\hfil##\hfil$\crcr#2\crcr\sim\crcr}}}
\def\simgt{\mathrel{\lower0.6ex\hbox{$\buildrel {\textstyle >}
 \over {\scriptstyle \sim}$}}}
\def\simlt{\mathrel{\lower0.6ex\hbox{$\buildrel {\textstyle <}
 \over {\scriptstyle \sim}$}}}

\input epsf

\newcommand{\hompc}{\,h\,{\rm Mpc}^{-1}}
\newcommand{\mpcoh}{\,h^{-1}\,{\rm Mpc}}
\newcommand{\ompc}{{\,\rm Mpc}^{-1}}
\newcommand{\mpc}{{\,\rm Mpc}}
\newcommand{\mnras}{MNRAS}

\newcommand{\aj}{AJL}

\def\be{\begin{equation}}
\def\ee{\end{equation}}
\def\ba{\begin{eqnarray}}
\def\ea{\end{eqnarray}}

\def\nn{\nonumber}

\newcommand{\bfk}{\mbox{\boldmath$k$}}
\newcommand{\bfp}{\mbox{\boldmath$p$}}
\newcommand{\bfq}{\mbox{\boldmath$q$}}
\newcommand{\sigmad}{\sigma_{\rm d}}

\begin{document}

\preprint{}

\title{Consistent Modified Gravity Analysis of Anisotropic Galaxy Clustering 
Using BOSS DR11}
 
\author{Yong-Seon Song$^{1}$, Atsushi Taruya$^{2,3}$, Eric Linder$^{4}$, Kazuya Koyama$^{5}$, Cristiano G. Sabiu$^{1}$, Gong-Bo Zhao$^{6,5}$, Francis Bernardeau$^{7,3}$, Takahiro Nishimichi$^{3,7,8}$, Teppei Okumura$^{3}$}
\email{ysong@kasi.re.kr}
\affiliation{$^{1}$ Korea Astronomy and Space Science Institute, Daejeon 305-348, R. Korea \\
$^2$Yukawa Institute for Theoretical Physics, Kyoto University, Kyoto 606-8502, Japan\\
$^{3}$Kavli Institute for the Physics and Mathematics of the Universe (WPI), The University of Tokyo Institutes for Advanced Study, The University of Tokyo, 5-1-5 Kashiwanoha, Kashiwa 277-8583, Japan\\
$^3$Kavli Institute for the Physics and Mathematics of the Universe, Todai Institutes for Advanced Study, the University of Tokyo, Kashiwa, Chiba 277-8583, Japan\\
$^4$Berkeley Lab and Berkeley Center for Cosmological Physics, 
University of California, Berkeley, CA 94720, USA\\
$^5$Institute of Cosmology $\&$ Gravitation, University of Portsmouth, Portsmouth, PO1 3FX, UK \\
$^6$National Astronomy Observatories, Chinese Academy of Science, Beijing, 100012, P. R. China\\
$^7$CNRS \& UPMC, UMR 7095, Institut d’Astrophysique de Paris, F-75014, Paris, France \\
$^{8}$CREST, JST, 4-1-8 Honcho, Kawaguchi, Saitama, 332-0012, Japan}
\date{\today}

\begin{abstract}
We analyse the clustering of cosmic large scale structure using a consistent 
modified gravity perturbation theory, accounting for anisotropic effects along 
and transverse to the line of sight. The growth factor has a particular scale 
dependence in $f(R)$ gravity and we fit for the shape parameter $f_{R0}$ 
simultaneously with the distance and the large scale (general relativity) limit 
of the growth function. Using more than 690,000 galaxies in the Baryon 
Oscillation Spectroscopy Survey Data Release 11, we find no 
evidence for extra scale dependence, with the 95\% confidence 
upper limit $|f_{R0}| <8 \times 10^{-4}$. Future clustering data, such as 
from the Dark Energy Spectroscopic Instrument, can use this consistent 
methodology to impose tighter constraints. 
\end{abstract}



\maketitle

\section{Introduction}

The growth of large scale structure in the universe is a multifaceted 
probe of cosmology. The galaxy clustering pattern measures cosmic 
geometry through baryon acoustic oscillations, giving angular distances 
transverse to the line of sight and radial distance intervals, or the 
Hubble parameter, along the line of sight, and the ratio of the two known 
as the Alcock-Paczy\'{n}ski effect~\cite{Alcock:1979mp}. The evolution of the clustering amplitude 
provides the growth factor and the shape of the clustering correlation 
function or power spectrum depends on early universe conditions and later, 
scale dependent effects. In addition, redshift space distortions (RSD) cause 
anisotropy in the clustering between the transverse and radial directions, 
and this probes the velocity field and the law of gravity~\cite{Kaiser:1987qv,Song:2008qt,Linder2007,Wang:2007ht,Nesseris:2007pa,White:2008jy}. 

General relativity (GR) predicts scale independent growth in the linear 
perturbation regime and specific redshift distortion patterns, 
and so probing for scale dependence or distortion deviations can test the 
theory of gravity. Our aim is to investigate the general relativity 
cosmological framework by fitting the galaxy clustering data while allowing 
for scale dependence, and constrain such deviations. We are particularly 
motivated by scalar-tensor theories and use a perturbation theory template 
derived for $f(R)$ gravity. Scale dependence arises at length scales smaller 
than or of order 
the inverse of the scalaron mass $m=1/\sqrt{3f_{RR}}$, where a subscript $R$ 
denotes a derivative with respect to the Ricci scalar~\cite{Song:2006ej}. For GR, $f=R$ and so 
the scale dependence vanishes. This scale dependence was tested using the combination of cosmic microwave background (CMB) data and galaxy clustering spectra in the following work~\cite{Song:2007da}.

By employing the perturbation theory of Taruya et al.~\cite{Taruya:2014faa}
that uses a resummed propagator to partially include nonperturbative and 
screening effects of the modified gravity, we can analyze the clustering 
correlation function to smaller scales than linear theory or 
simple perturbation calculations. We join this to our previous, 
substantially model independent approach of treating the background 
expansion in terms of the angular diameter distance and Hubble parameter
~\cite{Song13091162,2014PhRvD..89f3525L,2014JCAP...12..005S}, rather 
than assuming a dark energy model such 
as $\Lambda$CDM. Furthermore, we 
generalize the previous scale independent growth factor to two quantities: 
one scale independent (corresponding to the large scale limit of the growth 
factor) and one scale dependent (which can be thought of as characterizing 
the scalar-tensor modification, e.g.\ $m$ or $f_{R0}$). 

While we concentrate here on improving the RSD and galaxy clustering 
analysis for testing GR, various other methods have also been used. 
For example, the Planck collaboration has put constraints on a wide of range 
of modified gravity models using the Planck 2015 CMB data, combined with large scale structure observations \cite{MG:planck2015}; the SDSS-III (BOSS) team has tested GR using the observed structure growth patterns \cite{MG:boss,Sanchez:2013tga,Beutler:2013yhm}; the Wiggle-z team has tested modified gravity models \cite{Dossett:2014oia,Johnson:2015aaa}; and the CFHTLenS team employs the complementarity between weak 
gravitational lensing and RSD \cite{MG:cfhtlens}. The abundance of clusters \cite{MG:clusterc1,MG:clusterc2} and the cluster profiles \cite{MG:clusterprof1,MG:clusterprof2} have also been used for gravity tests. For more recent observational tests 
of GR, see \cite{MG:para1,MG:para2,MG:para3,Daniel1002,Daniel1008,MG:para4,MG:EFT,MG:astrotest} and \cite{Koyama:2015vza} for a review.  

To focus on exploring the scale dependence, we use only the clustering data, 
from the Baryon Oscillation Spectroscopic Survey (BOSS) Data Release 11 (DR11) 
of the Sloan Digital Sky Survey 3 (SDSS3)~\cite{Eisenstein:2011sa,Alam:2015mbd}. This consists of about 690,000 
galaxies over an effective volume of 6 Gpc$^3$ with an effective redshift 
$z=0.57$. We measure the two dimensional anisotropic correlation function 
as a function of transverse and radial separation between galaxies, and 
fit this to the redshift space distorted resummed perturbation theory. 
This comparison then imposes constraints on the cosmological and 
gravitational quantities, allowing us to test general relativity. 

Section~\ref{sec:all2} describes the modified gravity theoretical approach, 
from the $f(R)$ 
gravity model to the resummed propagator and resulting perturbation theory 
to the prediction for the clustering correlation function. We also lay out 
our approach of splitting the growth function into scale independent 
and dependent parts. In Sec.~\ref{sec:corrfn} we discuss measurement of 
the anisotropic correlation function from the data and treatment of the 
covariance matrix. We verify in Sec.~\ref{sec:testtempl} that we recover 
$\Lambda$CDM from $\Lambda$CDM simulated data. 
The comparison of the theory to the measurement is 
in Sec.~\ref{sec:cosmo1} and~\ref{sec:cosmo2}, where we analyze the fits 
to the cosmological quantities and their consistency with general 
relativity and $\Lambda$CDM. 
We summarize and conclude in Sec.~\ref{sec:concl}.

\section{theoretical model} \label{sec:all2} 

\subsection{$f(R)$ gravity model}\label{sec:fR1}

We consider perturbations around the Friedman-Robertson-Walker universe described by the metric 
\begin{equation}
ds^2 = -(1+ 2 \Psi) dt^2 + a(t)^2 (1 - 2 \Phi) \delta_{ij} dx^i dx^j.
\end{equation}
We will investigate modified gravity models that can be modelled by Brans-Dicke gravity on subhorizon scales. The metric perturbations and the scalar field perturbation $\phi = \phi_0 + \varphi$ obey the following equations in Fourier space \cite{Koyama:2009me}
\begin{eqnarray}
- k^2\Psi &=& 4 \pi G a^2 \rho \delta + \frac{1}{2} k^2 \varphi,
\label{poisson} \\
(3 +2 \omega_{\rm BD}) \frac{1}{a^2} k^2 \varphi
&=&  8 \pi G \rho_m \delta - {\cal I}(\varphi), 
\label{scalar}
\\
  \Phi - \Psi &=&  \varphi,  
\end{eqnarray}
where ${\cal I}$ represents the self-interaction, which can be expanded as  \cite{Koyama:2009me}
\begin{widetext}
\begin{eqnarray}
{\cal I} (\varphi)
&=& M_1(k) \varphi(k)  
   + \frac{1}{2} \int \frac{d^3 \bfk_1 d^3 \bfk_2}
{(2 \pi)^3} \delta_D(\bfk -\bfk_{12}) M_2(\bfk_1, \bfk_2)
\varphi(\bfk_1) \varphi(\bfk_2)  \nonumber\\
&& + \frac{1}{6}
\int \frac{d^3 \bfk_1 d^3 \bfk_2 d^3 \bfk_3}{(2 \pi)^6}
\delta_D(\bfk - \bfk_{123}) M_3(\bfk_1, \bfk_2, \bfk_3) 
\varphi(\bfk_1)\varphi(\bfk_2) \varphi(\bfk_3),
\end{eqnarray}
\end{widetext}
where $\bfk_{ij}=\bfk_i+\bfk_j$ and $\bfk_{ijk}=\bfk_i+\bfk_j+\bfk_k$.  
We treat the matter fluctuations $\delta$ as a pressureless fluid flow, whose evolution equation is given by 
\begin{eqnarray}
\frac{\partial \delta}{\partial t} + \frac{1}{a} \nabla \cdot 
[(1+ \delta) {\bf v} ] =0,
\label{continuity}
\\ 
\frac{\partial {\bf v}}{\partial t} + H {\bf v} 
+ \frac{1}{a} ({\bf v} \cdot \nabla) \cdot {\bf v} = - \frac{1}{a} \nabla \Psi.
\label{Euler}
\end{eqnarray}
We will assume the irrotationality of fluid quantities and express the velocity field in terms of the velocity divergence $\Theta = \nabla \cdot {\bf v}/(a H)$. 

\begin{figure*}
\begin{center}
\resizebox{3.4in}{!}{\includegraphics{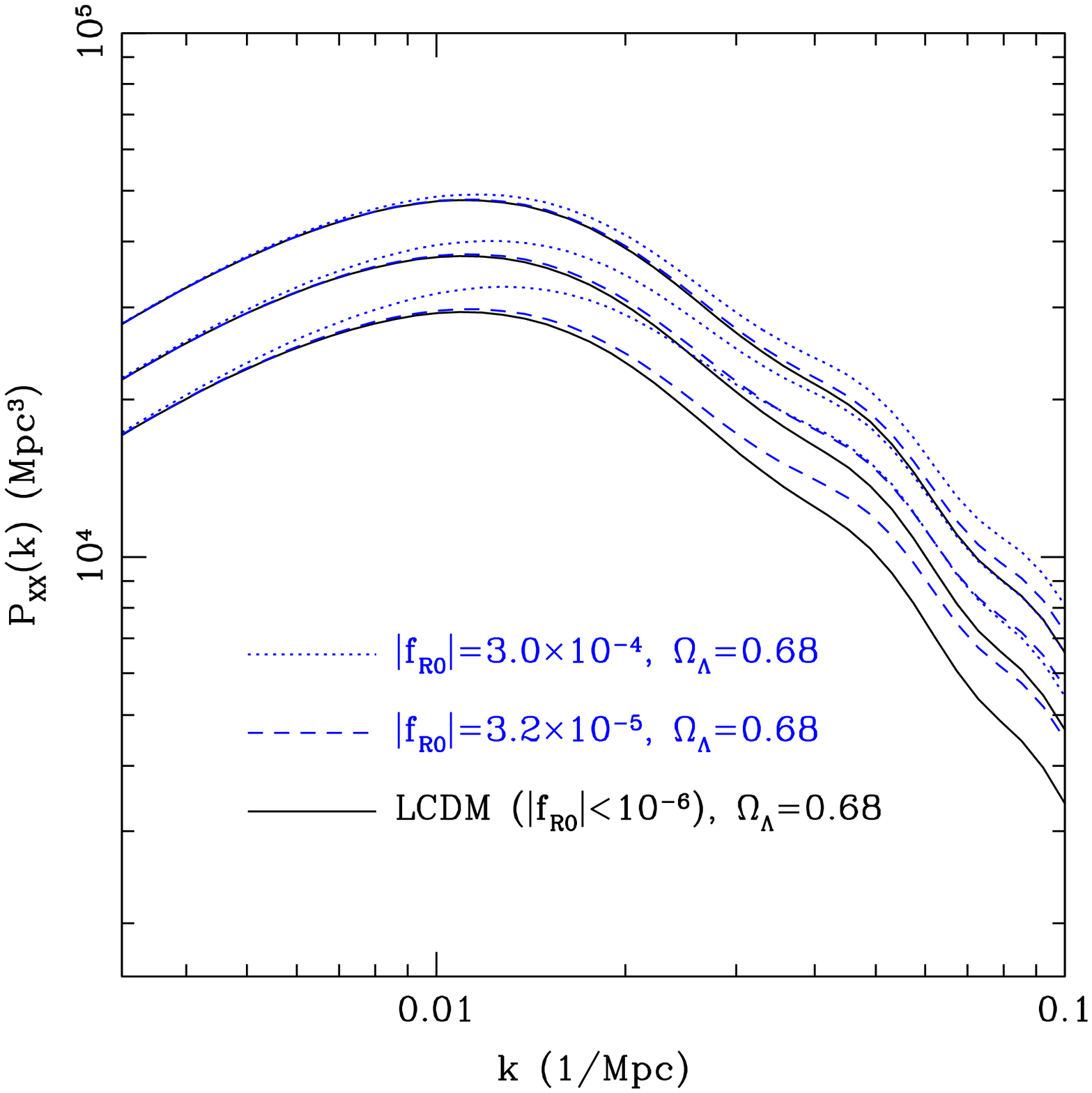}}
\resizebox{3.4in}{!}{\includegraphics{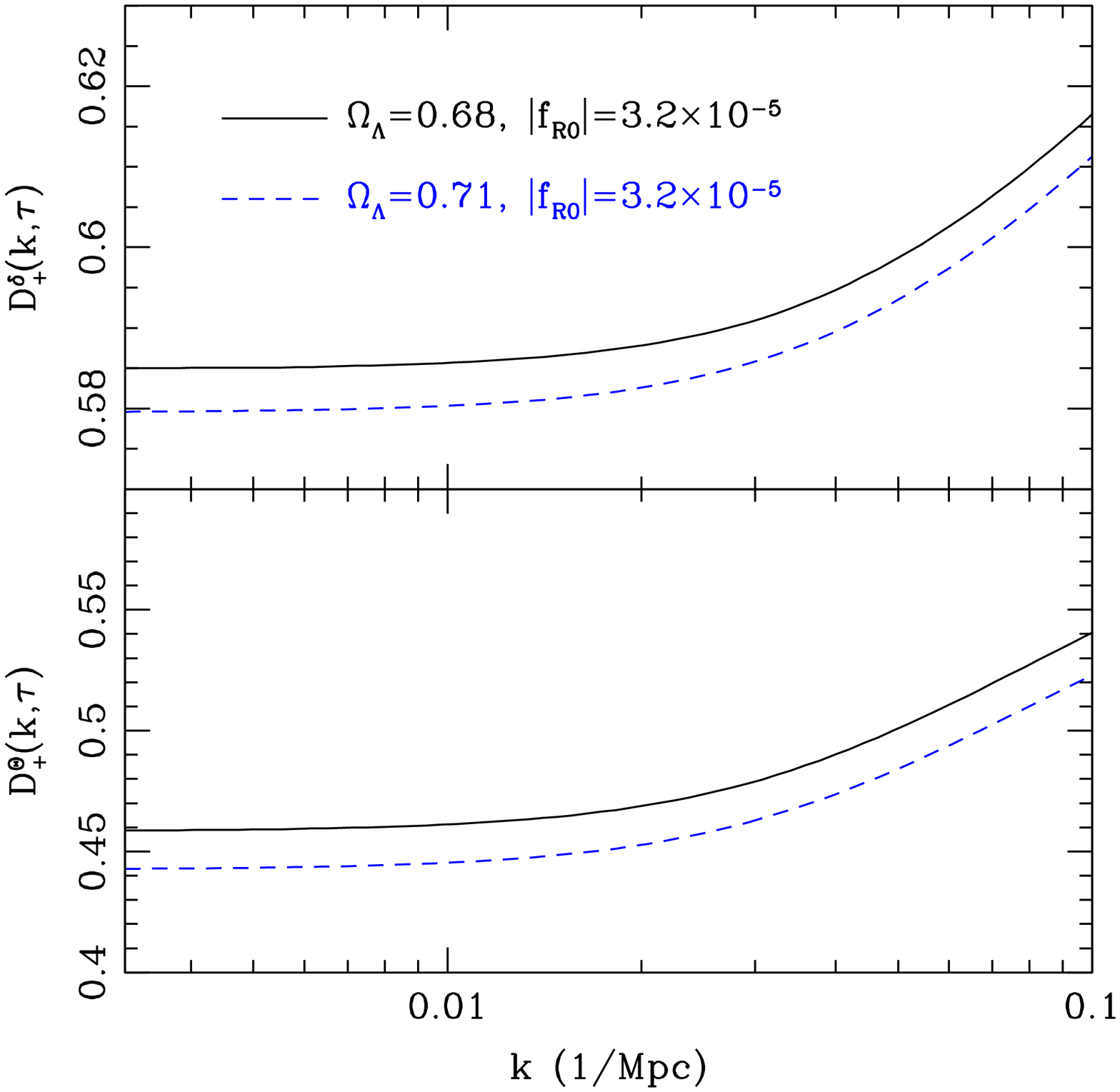}}
\end{center}
\caption{({\it Left panel}) The linear power spectra $P_{\delta\delta}(k)$, $P_{\delta\Theta}(k)$ and $P_{\Theta\Theta}(k)$ from top to bottom. The power spectra 
of the $\Lambda$CDM model are presented as black solid curves, and the power spectra of $f(R)$ gravity models with $|f_{R0}|=3.2\times 10^{-5}$ and $|f_{R0}|=3.0\times 10^{-4}$ are presented as blue dashed and dotted curves, respectively. 
The results are evaluated at redshift $z=0.57$. ({\it Right panel}) 
The growth functions $D_+^\delta$ and $D_+^\Theta$ have a scale independent 
amplitude, depending on $\Omega_\Lambda$ (taking fixed primordial amplitude), 
and scale dependent shape, depending on $|f_{R0}|$. The results in each 
subpanel are shown at $z=0.57$ for $|f_{R0}|=3.2\times 10^{-5}$, and 
$\Omega_\Lambda=0.68$ (black solid curves) and 0.71 (blue dashed curves). 
} 
\label{fig:Pk}
\end{figure*}

In order to develop the template for the redshift space power spectrum, we need to specify the interaction term ${\cal I}$. We consider $f(R)$ models as a representative class of models where the linear growth function is scale dependent. $f(R)$ gravity models are described by the action 
\begin{equation}
S= \int d^4 x \sqrt{-g} f(R) +  \int d^4 x \sqrt{-g} {\cal L}_m.
\end{equation}
We consider the function $f(R)$ given by the lowest order expansion in 
the small quantity $|f_{R0}| \ll 1$,  
\begin{equation}
f(R) = - 2 \kappa^2 \rho_{\Lambda} + |f_{R0}| \frac{\bar{R}_0^2}{R}, \label{eq:fRexp} 
\end{equation} 
general to many $f(R)$ models that are observationally viable. 
Here $\rho_{\Lambda}$ is the constant energy density and $\bar{R}_0$ is the background curvature at present time. For small $|f_{R0}|$, the background expansion 
in this Ansatz can be approximated as the one in the $\Lambda$CDM model. 

The scalar field perturbation is given by 
\begin{equation}
\varphi = f_R - \bar{f_R},
\end{equation}
where $f_R = df/dR$ and the bar indicates that the quantity is evaluated on the background. $f(R)$ gravity models are equivalent to Brans-Dicke gravity with $\omega_{\rm BD}=0$ and the interaction term ${\cal I}$ is given by 
\begin{equation}
{\cal I} = \delta R \equiv R(f_R) - R(\bar{f_R}).
\end{equation}
Thus the coupling functions $M_n$ are 
\begin{equation}
M_n = \frac{d^n \bar{R} (f_R)}{d f_R^n} \ , 
\end{equation} 
so Eq.~(\ref{eq:fRexp}) gives 
\begin{equation} 
M_1 = \frac{1}{2} \frac{1}{|f_{R0}|} \frac{\bar{R}^3}{\bar{R}_0^2} \ . 
\end{equation} 

By linearising the evolution equations, the solution for the gravitation 
potential is given by 
\begin{equation}
k^2 \Psi = - 4 \pi G \left( 
\frac{4 + M_1 a^2/k^2}{3 + M_1 a^2/k^2} \right) a^2 \rho_m \delta \ , 
\label{M1}
\end{equation}
On scales larger than the Compton wavelength of the scalar field, $m^{-1}=(3/M_1)^{1/2}=\sqrt{3 f_{RR}}$, the scalar field does not propagate and we recover $\Lambda$CDM. On the other hand, on small scales, gravity is enhanced: $G_{\rm eff}\rightarrow (4/3)G$. 

Combining this with the energy momentum conservation equations, we obtain the equation that determines the linear growth factor $D_+^\delta(k, t)$, 
\begin{equation}
{\cal L} D_+ ^\delta= 0, \quad {\cal L}= 
\frac{d^2}{d t^2} + 2 H \frac{d}{d t} - 4 \pi G \left( 
\frac{4 + M_1 a^2/k^2}{3 + M_1 a^2/k^2} \right) \rho_m. 
\label{eq:growth_factor}
\end{equation}
The growth function $D_+^\delta$ and the growth rate $D_+^\Theta \equiv d D_+^\delta/d \log a $ are also scale dependent. We introduce the following parametrisation of the growth rates
\ba
D_+^\delta(k, t) &=& G_{\delta}(t) F_\delta(k, t; M_1), \nn \\
D_+^\Theta(k, t) &=& G_{\Theta} (t) F_{\Theta}(k, t; M_1),
\label{eq:growth_rate}
\ea
where we defined $F_\delta(k, t; M_1)$ and $F_{\Theta}(k, t; M_1)$ so that $D_+^\delta(k, t) \to G_{\delta}(t)$ and $D_+^\Theta(k,t) \to G_{\Theta}(t)$ in the limit of $k \to 0$. Since we recover $\Lambda$CDM in the $k \to 0$ limit, $G_{\delta}$ and $G_{\Theta}$ are determined by the usual cosmological parameters and they are independent of $|f_{R0}|$. On the other hand, the scale dependence is controlled by $M_1$ in Eq.~(\ref{M1}), which is determined by $|f_{R0}|$ and the cosmological parameters.

We will find solutions for $\delta$ and $\theta$ by solving Eqs.~(\ref{poisson}), (\ref{scalar}), (\ref{continuity}), (\ref{Euler}) using perturbation theory. Once the nonlinearity becomes important, the non-linear self-interactions $M_{i>1}$ will suppress the scalar field interactions by the chameleon mechanism. This effect is included perturbatively in our approach.

\subsection{RSD model for $f(R)$ gravity models}\label{sec:fR2}
The anisotropy of galaxy clustering is now recognized as a useful probe of gravity on 
cosmological scales. This is because the anisotropic clustering signal contains information on both the  
cosmic expansion and growth of structure, 
through the Alcock-Paczynski effect and RSD. In principle, these two effects are simply described 
by the mapping formula from the statistically isotropic frame, however 
modeling the RSD effect is rather 
complex because of the nonlinear and stochastic nature of the mapping. 
As a result, the applicable range of the linear theory prediction is 
quite limited. Even at the largest scales accessible by future galaxy surveys, 
a proper account of the nonlinearity is crucial for a robust test of gravity.

Here, we will adopt an improved model of RSD by Ref.~\citep{Taruya:2010mx} 
for the theoretical template of the redshift-space correlation function. 
While this model has been originally proposed to characterize the matter power spectrum in GR, the assumptions and propositions behind the model prescription do not rely on any specific gravitational theory. 
Thus it can apply to any model of modified gravity.   
Indeed, the model has been tested against the dark matter simulation of the $f(R)$ gravity model, where a good agreement with $N$-body results was found \citep{Taruya:2013quf}.  One important remark is that 
the dynamics of density and velocity fields in modified gravity 
can be different from GR, and each building block in the RSD model
needs to be carefully computed, taking a proper account of the modification of 
(non)linear gravitational growth, which we will describe below 
(see also Appendix \ref{appendix:regpt}). 
This is what we refer to as a consistent analysis of modified gravity. 

Employing the linear bias prescription, the improved model of RSD is given 
in Fourier space as function of wavenumber $k$ and 
directional cosine $\mu=k_z/k$ with $k_z$ being line-of-sight 
component of $k$: 
\ba
\tilde{P}(k,\mu) &=& \left\{
b^2P_{\delta\delta}(k) + 2\mu^2 bG_\Theta P_{\delta\Theta}(k) 
+ \mu^4 G_\Theta^2P_{\Theta\Theta}(k)
\right.
\nonumber
\\
&&\left.
 + 
A(k,\mu;b,G_\Theta) + B(k,\mu;b,G_\Theta)\right\}\nn\\
&&\times D_{\rm FoG}(k\mu\sigma_p),
\label{eq:TNS10}
\ea
where $b$ is the linear bias parameter. (See Sec.~\ref{sec:cosmo1} for a 
more sophisticated treatment.) 
The $G_\Theta$ is the parameter characterizing the growth of structure 
introduced in Eq.~(\ref{eq:growth_rate}). The functions 
$P_{\delta\delta}$, $P_{\Theta\Theta}$ and $P_{\delta\Theta}$ are, 
respectively, the auto-power spectra of density and velocity-divergence fields, and their cross-power spectrum. 
The 
function $D_{\rm FoG}$ characterizes the suppression of the power spectrum 
due to the virialized random motion of galaxies~\cite{1972MNRAS.156P...1J,1999ApJ...517..531S,Scoccimarro:2004tg}, for which we assume the Gaussian form:
\be
D_{\rm FoG}(x)=e^{-x^2}.
\label{eq:Gaussian_FoG}
\ee
Since the suppression of the power spectrum basically comes from the galaxies sitting in a halo, we shall treat $\sigma_p$ in the damping function $D_{\rm FoG}$ as 
a free parameter. 

In Eq.~(\ref{eq:TNS10}), the main characteristic is 
the $A(k,\mu)$ and $B(k,\mu)$ terms, which represent the higher-order coupling between density and velocity fields. These have been derived on the basis of the low-$k$ expansion from the exact expression for the redshift-space power spectrum, expressed as:
\ba
A&=&b^3\sum_{n=1}^3\sum_{a,b=1}^2\mu^{2n}\left(\frac{G_{\Theta}}{b}\right)^{a+b-1}
\frac{k^3}{(2\pi)^2}\int_0^\infty dr\int_{-1}^1 dx
\nonumber\\
&&\times\Bigl\{
A^n_{ab}(r,x)\,B_{2ab}(\bfp,\bfk-\bfp,-\bfk)
\nonumber\\
&&\qquad+\widetilde{A}^n_{ab}(r,x)B_{2ab}(\bfk-\bfp,\bfp,-\bfk)\Bigr\},
\label{eq:Aterm}
\\
B&=&b^4\sum_{n=1}^4\sum_{a,b=1}^2\mu^{2n}\left(-\frac{G_{\Theta}}{b}\right)^{a+b}
\frac{k^3}{(2\pi)^2}\int_0^\infty dr\int_{-1}^1 dx
\nonumber\\
&&\times
\,B^n_{ab}(r,x)\frac{P_{a2}(k\sqrt{1+r^2-2rx})P_{b2}(kr)}{(1+r^2-2rx)^a}.
\label{eq:Bterm}
\ea
where $r=p/k$ and $x=\bfk\cdot\bfp/(k\,p)$. Here, the 
functions $P_{ab}$ and $B_{abc}$ are the power spectrum and bispectrum of the two-component multiplet $\Psi_a=(\delta,\Theta)$. The non-vanishing coefficients, $A^n_{ab}$, $\widetilde{A}^n_{ab}$, and $B^n_{ab}$,  are those presented in Sec.~III-B2 of Ref.~\citep{Taruya:2013my} and Appendix B of Ref.~\citep{Taruya:2010mx}, respectively.

In calculation of the redshift-space power spectrum, we need to properly take into account the effect of nonlinear gravitational evolution in each term of Eq.~(\ref{eq:TNS10}). Since the standard perturbation theory (PT) \citep{Bernardeau_review} is known to produce an ill-behaved expansion leading to unwanted UV behaviour, we shall apply the resummed perturbation theory called {\tt RegPT}, which has been formulated in Ref.~\citep{Taruya:2012ut} and been later extended in Ref.~\cite{Taruya:2014faa} to the modified gravity models. 
Following the prescription described in~\cite{Taruya:2014faa}, we compute the power spectra $P_{XY}(k)$ as well as the $A$ and $B$ terms, including consistently the nonlinear corrections up to the one-loop order in the $f(R)$ gravity model. The explicit expressions for statistical quantities, necessary for a consistent one-loop calculation of the redshift-space power spectrum, are summarized in Appendix \ref{appendix:regpt}.

\begin{figure*}
\begin{center}
\resizebox{3.4in}{!}{\includegraphics{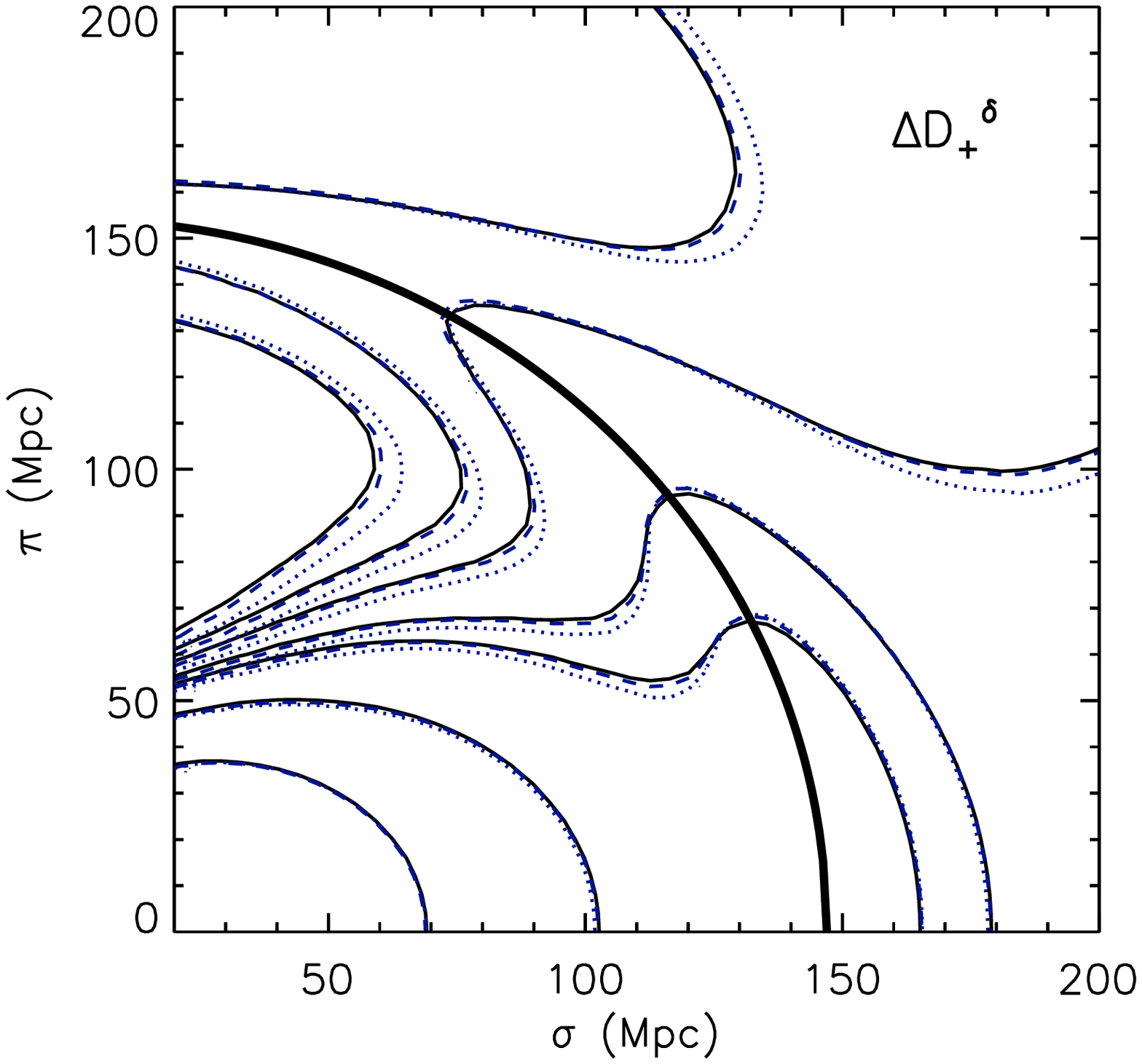}} 
\resizebox{3.4in}{!}{\includegraphics{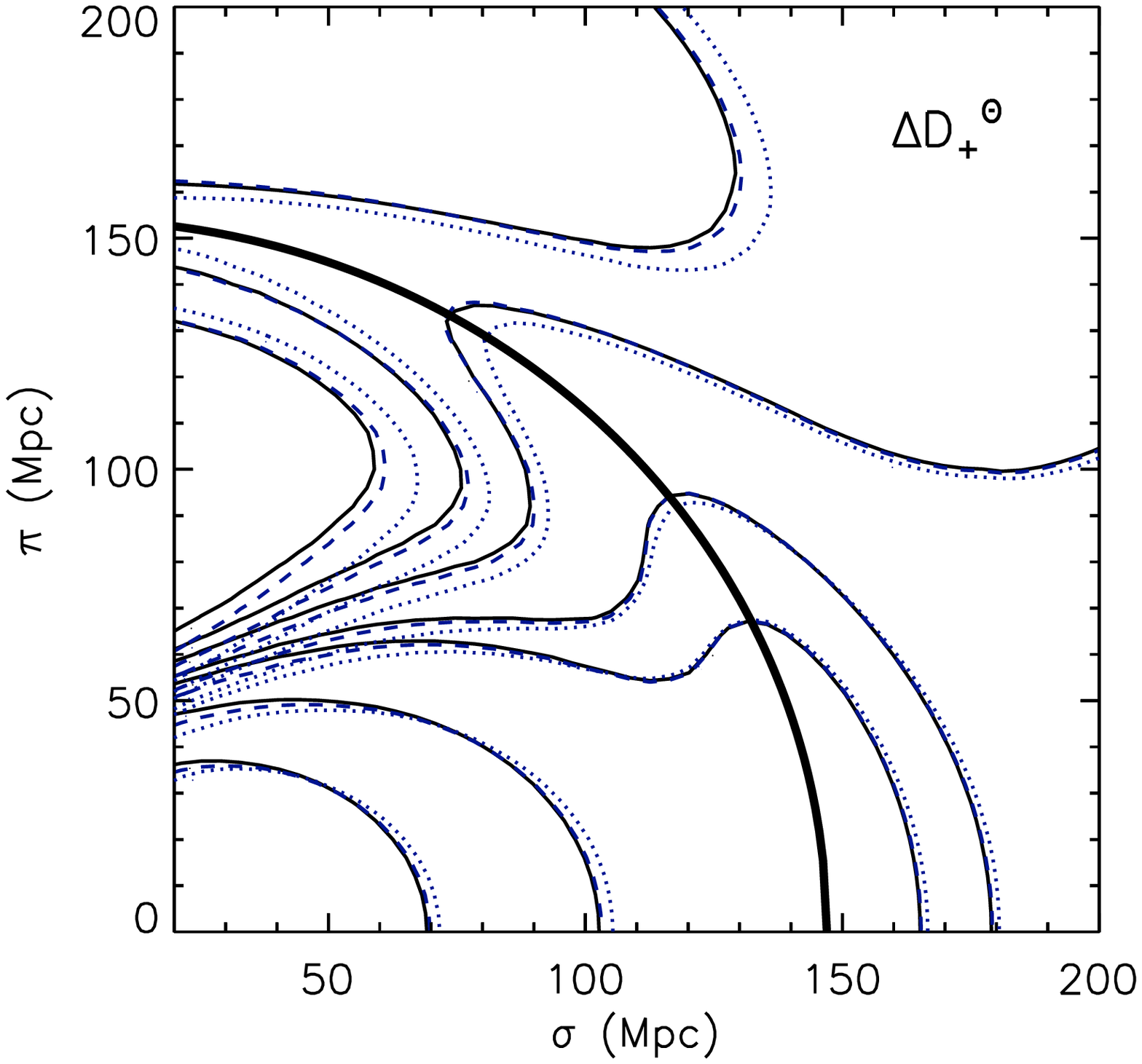}} 
\end{center}
\caption{({\it Left panel}) The correlation function $\xi(\sigma,\pi)$ of the 
$\Lambda$CDM model is shown as black solid contours. Blue dashed and dotted contours represent $\xi(\sigma,\pi)$ of the $\Lambda$CDM model in which $D_+^\delta$ is replaced by that of $f(R)$ gravity models with $|f_{R0}|=3.2\times 10^{-5}$ and $|f_{R0}|=3.0\times 10^{-4}$, respectively. The thick solid circle represents the BAO ring. The levels of contours are given by $(-0.005,-0.0025,-0.004,0.003,0.005,0.016,0.05)$ from outer to inner contours. ({\it Right panel}) Black solid contours represent $\xi(\sigma,\pi)$ of the $\Lambda$CDM model, and blue dashed and dotted contours represent $\xi(\sigma,\pi)$ of the $\Lambda$CDM model in which $D_+^\Theta$ is replaced by that of $f(R)$ gravity models of $|f_{R0}|=3.2\times 10^{-5}$ and $|f_{R0}|=3.0\times 10^{-4}$, respectively.}
\label{fig:xivariation_test}
\end{figure*}

In the cosmological analysis in Sec.~\ref{sec:fR4}, we will consider the following two cases, which can be thought of as fixing the expansion and large scale growth histories, or allowing them to vary. In the first case, we will fix all the cosmological parameters in the template by a fiducial cosmology and vary $|f_{R0}|$, the linear bias $b$ and the velocity dispersion $\sigma_p$. This corresponds to fixing $G_{\delta}$,  $G_{\Theta}$ and the amplitude of the initial spectrum to those determined by the fiducial cosmology.

In the second case, we allow $G_{\Theta}$ to float by varying $\Omega_{\Lambda} \equiv \rho_{\Lambda} /3 H_0^2$ where $H_0$ is the present day Hubble constant in the flat universe. 
Note that with the variation of $\Omega_\Lambda$, 
the amplitude of the growth function $D_+^{\delta}$ ($G_{\delta}$) as well as the shape of the growth function ($F_{\delta}$) and growth rate ($F_{\Theta}$) are also changed through the mass parameter $M_1$ [see Eq.~(\ref{eq:growth_factor})]. 
In practice, however, we see from Fig.~\ref{fig:Pk} that the dependence of the shapes $F_\delta$ and $F_{\Theta}$ on $\Omega_\Lambda$ are very weak. Further, the change of $G_{\delta}$ is degenerate with the linear bias $b$ in the linear regime. Although the one-loop terms break the degeneracy, this effect is small in the regime of our interest. Hence, in the second case, 
we vary the scale independent growth rate $G_{\Theta}$ and the linear bias 
in the PT template, and introduce a new parameter 
\begin{equation}
G_b(t) \equiv b\; G_\delta (t), 
\end{equation}
to represent the combined effect of the variation of $b$ and $G_{\delta}$. 
Then $G_{\delta}$ and the initial amplitude of the power spectrum are formally 
held fixed. 

\subsection{Correlation function $\xi{(\sigma,\pi)}$ in $f(R)$ gravity}\label{sec:fR3} 

The two-point correlation function of galaxy clustering in the redshift space, $\xi$, is described by a function of $\sigma$ and $\pi$, where $\sigma$ and $\pi$ are the transverse and the radial directions with respect to the observer. From the power spectrum $\tilde{P}(k, \mu)$, we can compute the correlation function $\xi(\sigma,\pi)$ by Fourier transformation. The correlation function is generally expanded as
\ba\label{eq:xi_eq}
\xi(\sigma,\pi)&=&\int \frac{d^3k}{(2\pi)^3} \tilde{P}(k,\mu)e^{i{\bf k}\cdot{\bf s}}\nn\\
&=&\sum_{\ell:{\rm even}}\xi_\ell(s) {\cal P}_\ell(\nu)\,,
\label{moments}
\ea
with ${\cal P}_{\ell}$ being the Legendre polynomials. Here, we defined $\nu=\pi/s$ and $s=(\sigma^2+\pi^2)^{1/2}$. The moments of the correlation function, $\xi_\ell(s)$, are defined in~\cite{Song13091162}.
The contributions from moments higher than $\ell=8$ will be ignored in our analysis because we are only interested in quasi-nonlinear scales where perturbation theory is applicable. 


We compute theoretical templates for the moments $\xi_\ell(s)$ using the power spectrum given by Eq.~(\ref{eq:TNS10}). The shape of the initial spectra and the growth functions, $G_\delta$ and $G_\Theta$, are given by the best fit $\Lambda$CDM model from Planck 2013 \cite{Ade:2013zuv}.
We provide multiple templates with various $|f_{R0}|=(0,10^{-6},3.2\times 10^{-6},10^{-5},3.2\times 10^{-5},10^{-4},3.2\times 10^{-4},10^{-3},3.2\times 10^{-3})$. For other values of $|f_{R0}|$, we interpolate these templates. 


We now study the variation of $\xi(\sigma,\pi)$ due to the change of the growth function and growth rate. This variation was studied in the case of scale independent growth functions in \cite{Song13091162} by varying $G_{\delta}(t)$ and $G_{\Theta}(t)$ in the {\tt RegPT} template for $\Lambda$CDM. Following this approach, we first study the impact of having a scale dependent growth function or growth rate. For this purpose, we compute the correlation function $\xi(\sigma, \pi)$ using a $\Lambda$CDM template and replace the growth function $D_+^{\delta}$ or growth rate $D_+^{\Theta}$ by that in $f(R)$ gravity with $|f_{R0}|=3.2\times 10^{-5}$ and $|f_{R0}|=3.0\times 10^{-4}$. 

For the scale dependent growth function $D_+^\delta$, the variation of $\xi(\sigma,\pi)$ with a small $|f_{R0}|=3.2\times 10^{-5}$ is similar to the case of a scale independent enhancement of the growth function studied in \cite{Song13091162}. Peak points on the BAO ring represented by a thick black solid curve in Fig.~\ref{fig:xivariation_test} move coherently along the circle in an anti--clockwise direction. The blue dashed contours in the left panel of Fig.\ref{fig:xivariation_test} represent this variation. However,  $\xi(\sigma,\pi)$ with a larger $|f_{R0}|=3.0\times 10^{-4}$ varies differently from the scale independent case. Peak points on the BAO ring remain the same, while minima of BAO are deepened, shown as blue dotted contours in the same panel. 

Next, we consider the variation of $\xi(\sigma,\pi)$ due to the scale dependent growth rate $D_+^\Theta$. In the case of the scale independent growth rate, if $G_\Theta$ increases or decreases, the anisotropic effects from higher order moments are visible in the plot of $\xi(\sigma,\pi)$. The location of the crossing points between the contour levels and the BAO ring (thick solid) shifts clockwise or anti-clockwise slightly.
The blue dashed contours in the right panel of Fig.~\ref{fig:xivariation_test} represent the variation of $\xi(\sigma,\pi)$ with $\Delta D_+^\Theta$ for $|f_{R0}|=3.2\times 10^{-5}$ 
and $|f_{R0}|=3.0\times 10^{-4}$. For $|f_{R0}|=3.0\times 10^{-4}$, we can see that the peak positions are `squeezed' along the BAO ring. 

\begin{figure}
\begin{center}
\resizebox{3.4in}{!}{\includegraphics{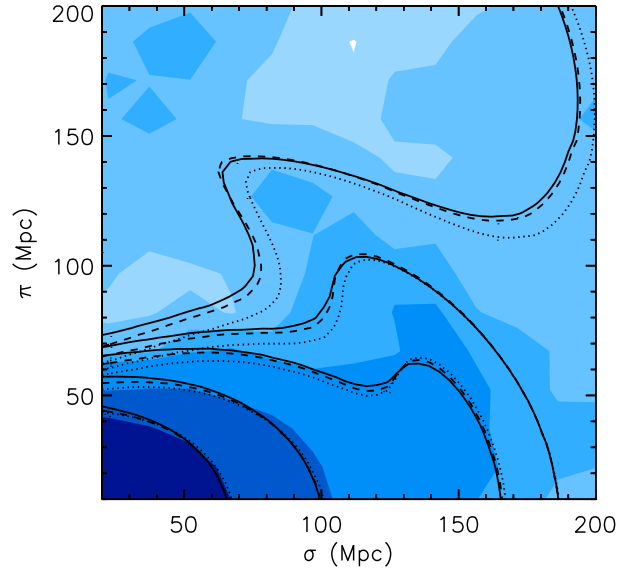}}
\end{center}
\caption{The best fit correlation function $\xi(\sigma,\pi)$ of $\Lambda$CDM (black solid unfilled contours) and the correlation function of $f(R)$ gravity models with $|f_{R0}|=3.2\times 10^{-5}$ (black dashed unfilled contours) and $3.0\times 10^{-4}$ (dotted unfilled contours). The blue filled contours represent the measured $\xi(\sigma,\pi)$ from the DR11 CMASS data. The levels of contours are given by $(-0.001,0.002,0.005,0.016,0.05)$ from the outer to inner contours.} 
\label{fig:DR11}
\end{figure}

Having shown the individual effects of a scale dependent growth function and growth rate on the correlation function, we now present the correlation function $\xi(\sigma,\pi)$ in $f(R)$ gravity models. 
In Fig.~\ref{fig:DR11}, the correlation function with $|f_{R0}|=3.2\times 10^{-5}$ and $|f_{R0}|=3.0\times 10^{-4}$ are plotted as black dashed and black dotted contours, respectively. There is no variation of $\xi(\sigma,\pi)$ up to $|f_{R0}|\la 10^{-6}$, and the correlation function is effectively equivalent to that of $\Lambda$CDM. When $|f_{R0}|$ increases to $|f_{R0}|\sim 10^{-4}$, we observe the deviation of $\xi(\sigma,\pi)$ from $\Lambda$CDM and this deviation can be understood as the combined effect of the scale dependent growth function and growth rate shown in Fig.~\ref{fig:xivariation_test}.  

\begin{figure*}
\begin{center}
\resizebox{3.4in}{!}{\includegraphics{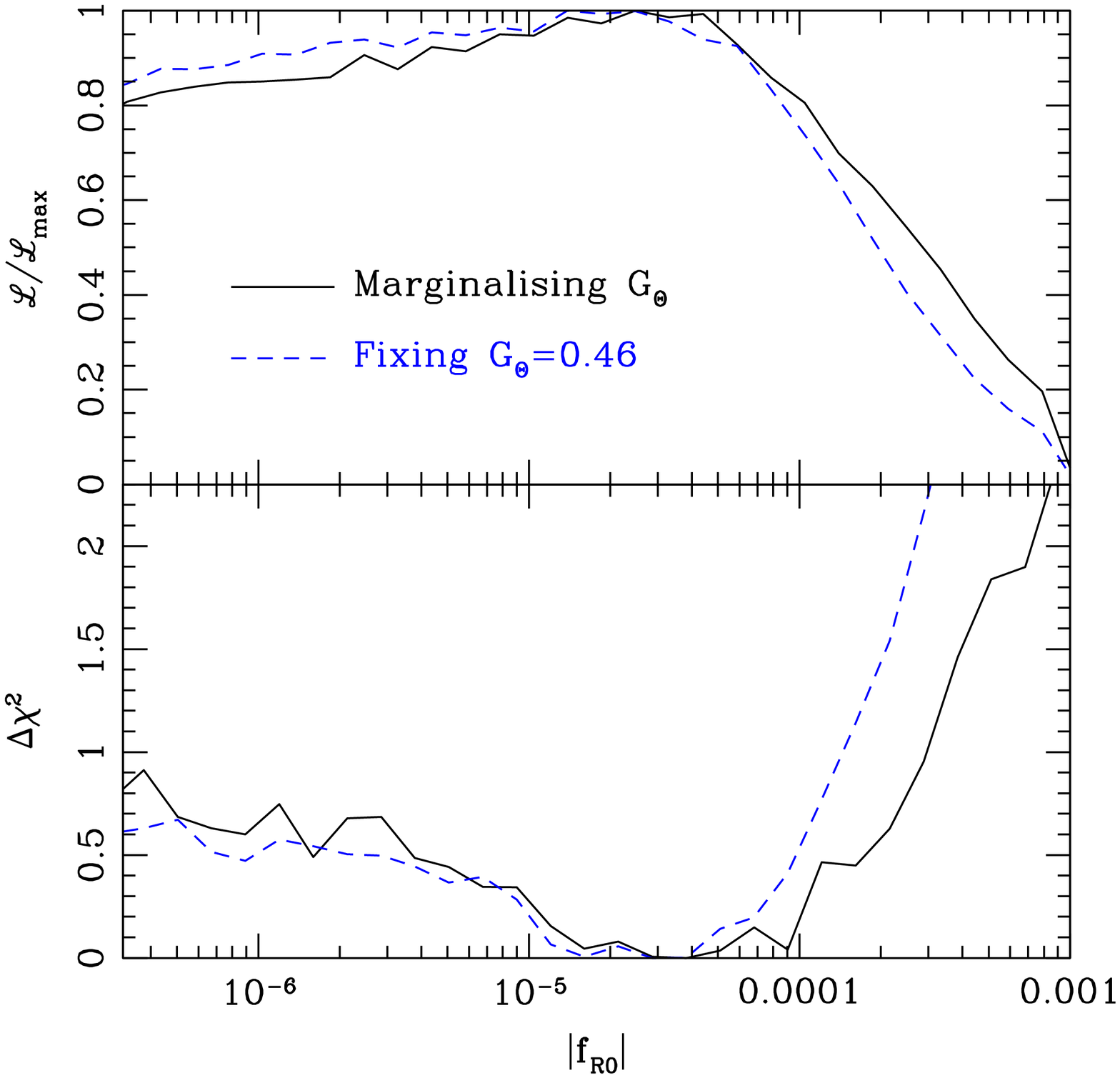}}
\resizebox{3.4in}{!}{\includegraphics{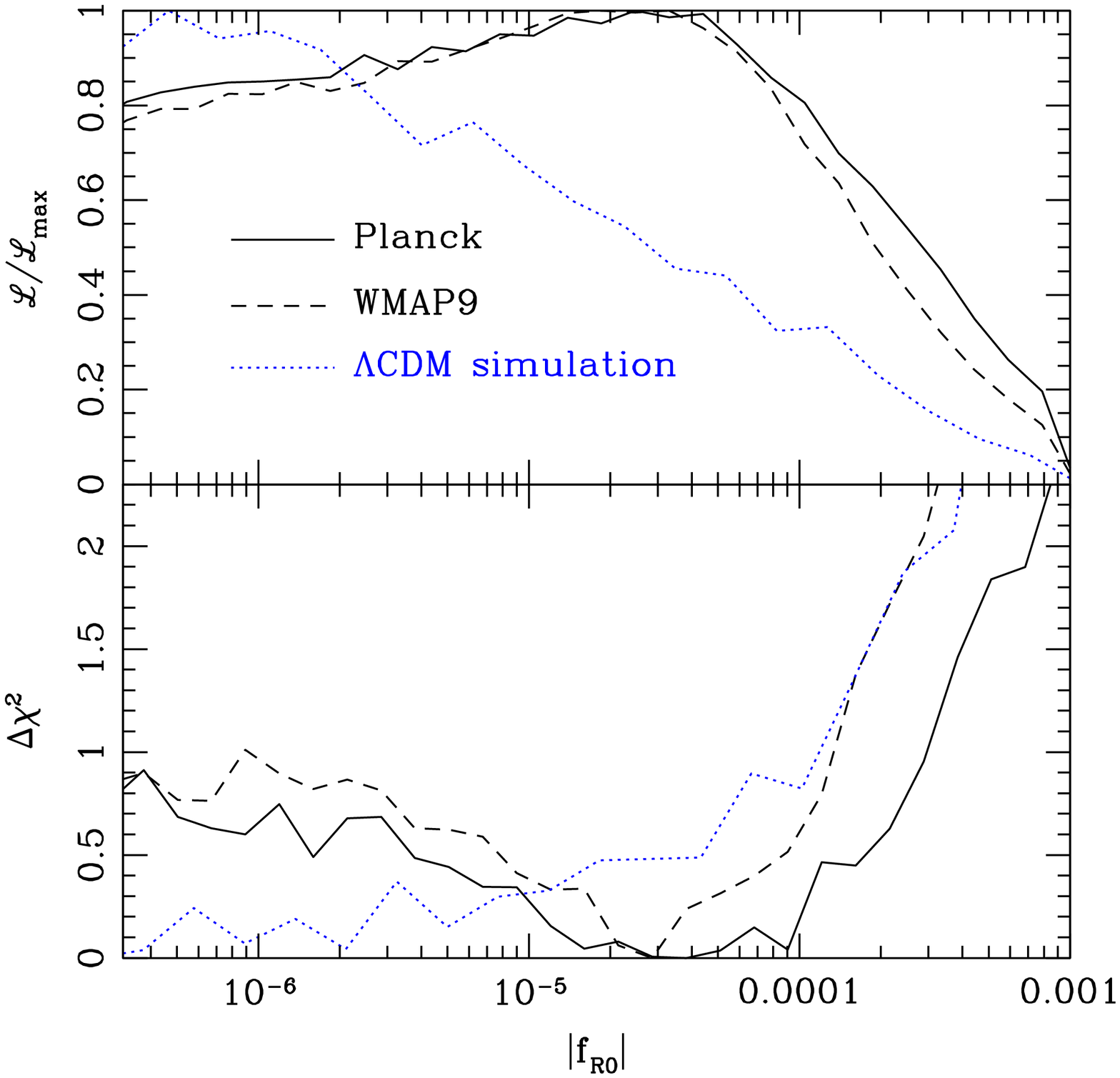}}
\end{center}
\caption{The measured constraints on $f_{R0}$, and their robustness to 
various tests, are presented. The measured likelihood function appears in 
the top panels and the measured difference of $\chi^2$ is in the bottom panels. 
{\it (Left panel)\/} Results marginalizing over the scale independent growth 
rate $G_\Theta$ are shown by the black solid curve, while the constraints 
fixing $G_{\Theta}=0.46$, given by the Planck concordance $\Lambda$CDM model, 
are blue dashed curves. The results for $f_{R0}$ do not depend appreciably 
on the scale independent behavior. {\it (Right panel)} The results also do 
not depend significantly on whether the initial power spectrum $P(k)$ used 
matches the Planck (black solid) or WMAP9 (black dashed) model. The blue 
dotted curve represents the results from analyzing galaxy clustering from 
$\Lambda$CDM mock catalogues, verifying that $|f_{R0}|\to0$ is recovered 
in this case.}
\label{fig:measuredfR}
\end{figure*}

\section{methodology and results}\label{sec:fR4}

The observed clustering of galaxies in redshift space not only probes the density and velocity fields, i.e.\ the growth and gravity as discussed in the previous 
section, but also provides a useful tool to determine both the transverse and radial distances by  exploiting the Alcock--Paczy\'nski effect and the BAO scale. In galaxy redshift surveys, each galaxy is located by its angular coordinates and redshift. However, the correlation function, $\xi(\sigma,\pi)$, is measured in comoving distances. Therefore a fiducial cosmological model is required for conversion into comoving space. We use the best fit $\Lambda$CDM universe to Planck 2013 data. 
The conversion depends on the transverse and radial distances involving $D_A$ and $H^{-1}$. Instead of recreating the measured correlation function in comoving distances for each different model, we create the fiducial maps from the theoretical correlation function by rescaling the transverse and radial distances using $D_A$ and $H^{-1}$ and fit them to the observed correlation function. Therefore, when we fit the measured $\xi(\sigma,\pi)$, the two distance parameters of ($D_A$, $H^{-1}$) are added to the structure formation parameter set of $\{G_{\delta},G_{\Theta},\sigma_p,|f_{R0}|,\sigma_p\}$ discussed in Sec.~\ref{sec:fR2}. 

\subsection{Measured $\xi(\sigma,\pi)$ using DR11}\label{sec:corrfn} 

Our measurements are based on those previously presented in \citep{2014JCAP...12..005S} which follows a similar procedure to  \citep{2014PhRvD..89f3525L}.

Briefly, in our analysis we utilise data release DR11 of the Baryon Oscillation Spectroscopic Survey \citep[BOSS; ][]{2012AJ....144..144B, 2013AJ....145...10D, 2013AJ....146...32S} which is part of the larger Sloan Digital Sky Survey \citep[SDSS; ][]{2000AJ....120.1579Y, 2006AJ....131.2332G} program. From DR11 we focus our analysis on the {\em Constant Stellar Mass Sample} (CMASS) \cite{CMASS}, which contains  690,826 galaxies  and covers the redshift range $z=0.43-0.7$ over a sky area of $\sim$8,500 square degrees with an effective volume of  $V_{\rm eff}\sim 6.0\,$Gpc$^3$. The CMASS galaxy sample is composed primarily of bright, central galaxies, resulting in a highly biased ($b\sim2$) selection of mass tracers \citep{2013MNRAS.432..743N}. 

The redshift-space two-dimensional correlation function 
$\xi(\sigma,\pi)$ of the BOSS DR11 galaxies was computed  
using the standard Landy-Szalay estimator~\cite{1993ApJ...412...64L}. 
In the computation of this estimator we used a random point catalogue that constitutes an unclustered but observationally representative sample of the BOSS CMASS survey and  contains $\sim50$ times as many randoms as we have galaxies. 

The covariance matrix was obtained from 600 mock catalogues based on second-order Lagrangian perturbation theory (2LPT)  \cite{Manera:2012sc, Reid:2012sw}. The mocks reproduce the same survey geometry and number density as the CMASS galaxy sample. We obtain the covariance matrix using the same treatment presented in our previous works \cite{2014PhRvD..89f3525L,2014JCAP...12..005S}.


We calculate the correlation function in 225 bins spaced by $10\mpcoh$ in the range $0<\sigma,\pi<150\mpcoh$. However, at small scales, if the non--perturbative effect of FoG is underestimated, 
then the residual squeezing can be misinterpreted as a variation in $G_{\theta}$ or indeed $f_{R0}.$
We expect the FoG effect to be increasingly important at smaller scales, 
and so these measurements may be at risk of misestimation. We therefore impose a conservative cut on the measurements, 
excluding  $\sigma_{\rm cut} < 40\mpcoh$ and  $s_{\rm cut} < 50\mpcoh$~\cite{2014PhRvD..89f3525L}. 
Indeed, ~\cite{2014PhRvD..89f3525L} showed that cosmological parameter bias began to occur at smaller scales. This reduces the number of measurement bins in $\sigma$ and $\pi$ to $N_{bins}=163$.

\subsection{Tests of theoretical templates} \label{sec:testtempl} 

When the conservative cut--off scales of $\sigma_{\rm cut} = 40\mpcoh$ and  $s_{\rm cut} = 50\mpcoh$ are used for the analysis, the effective range of scale in Fourier space becomes $k<0.1\ompc$. The power spectra of $\Lambda$CDM and $f(R)$ gravity models are presented in this range of scale in Fig.~\ref{fig:Pk}. 
For the clustering scales considered in this likelihood analysis, there are no deviations from $\Lambda$CDM. This implies that $f(R)$ gravity models with $\log |f_{R0}|\la -6$ are effectively equivalent to $\Lambda$CDM in this analysis. 
We take a uniform prior on $\log |f_{R0}|$ between $-7$ and $-3$.

We first test our pipeline of analysis by checking whether it is possible to recover the $\Lambda$CDM limit $\log |f_{R0}|\la -6$ 
using the mock catalogues based on $\Lambda$CDM. 
We use the 611 CMASS mock catalogues to measure central values of $\xi(\sigma,\pi)$ and fit our theoretical $f(R)$ templates to the observed correlation function. The measured likelihood function of $\log |f_{R0}|$ is presented as a blue dotted curve in the right panel of Fig.~\ref{fig:measuredfR}. 
The best fit $\log |f_{R0}|$ indeed lies within the $\Lambda$CDM limit of log $|f_{R0}|\la -6$. 
There are no mock galaxy catalogues based on $f(R)$ gravity available so we are not able to fully test our theoretical templates away from the $\Lambda$CDM limit. The perturbation theory predictions for the redshift space power spectrum in Fourier 
space were, however, tested against N-body simulations for $f(R)$ gravity models and it was shown that the perturbation theory based template was able to reproduce the input value of $\log |f_{R0}|$ of the simulation in an unbiased way for $k <0.1$ Mpc$^{-1}$ \cite{Taruya:2013quf}.

\subsection{Constraints on $f(R)$ gravity}\label{sec:cosmo1} 

We now present the results for constraints on $|f_{R0}|$, summarized in 
Table~\ref{tab:measurements}. The Markov Chain Monte Carlo (MCMC) method is used to sample a probability distribution. The normalised distribution function of ${\cal L}/{\cal L}_{\rm max}$ of the chain is given in terms of $\log |f_{R0}|$ in the top panel of Fig.~\ref{fig:measuredfR}, and the $\Delta\chi^2$ is shown in the bottom panel.

We begin by analysing our results while fixing $G_\Theta$ to the Planck $\Lambda$CDM model prediction of $G_\Theta=0.46$. The measured $\log |f_{R0}|$ is then 
$\log |f_{R0}|<-3.87$ and $\log |f_{R0}|<-3.6$ at 68\% and 95\% confidence upper limits respectively. The lower index denotes the prior is hit -- 
recall the $\Lambda$CDM limit of $\log |f_{R0}|\la -6$, and the upper indices denote 68\% and 95\% confidence levels of $\log |f_{R0}|$ represented by unbracketed and bracketed numbers respectively. The distribution function and $\Delta\chi^2$ are presented by blue dashed curve in the left panel of Fig.~\ref{fig:measuredfR}. 

The measured $\log f_{R0}$ after marginalising $G_\Theta$ and all other parameters is $-4.5_{\rm [prior]}^{+0.96(1.4)}$ as presented in Table~\ref{tab:measurements}. The best fit $|f_{R0}|$ is $3.2\times 10^{-5}$, and the upper bounds are $|f_{R0}|=3.0\times 10^{-4}$ and $8.0\times 10^{-4}$ at 68\% and 95\% confidence levels, respectively. The observed $\xi(\sigma,\pi)$ is presented as blue filled contours in Fig~\ref{fig:DR11}, and $\xi(\sigma,\pi)$ of the best fit $f(R)$ gravity model is represented as black dashed contours, which show a slightly better fit than $\xi(\sigma,\pi)$ of $\Lambda$CDM represented as black solid contours. 
The black solid curves in the left panel of Fig.~\ref{fig:measuredfR} represents the distribution function (upper) and $\Delta\chi^2$ (lower). 
Note that $\log |f_{R0}|<-6$, i.e.  $\Lambda$CDM, is within $\Delta\chi^2<1$ of our best fit value and so is wholly allowed by the current data. When $|f_{R0}|$ is larger than $10^{-4}$, $\xi(\sigma,\pi)$ becomes significantly different from the $\Lambda$CDM prediction. The black dotted contours in Fig.~\ref{fig:DR11} represent $\xi(\sigma,\pi)$ with $|f_{R0}|=3.0\times 10^{-4}$ at 68\% confidence bound. 
The difference from the measured $\xi(\sigma,\pi)$ can be observed. 


\label{sec:minprior}
\begin{table*}
\renewcommand{\arraystretch}{1.45} 
\setlength{\tabcolsep}{0.2cm}
\begin{center}
\begin{tabular}{|l|c|c|c|c|}
\hline
Parameters & Planck & Measurements & Fixed $G_{\Theta}=0.46$ & Fixed $|f_{R0}|=0$\\
\hline
\hline
$D_A\,(\mpc)$       & $1397.5$ & $1428.0_{-28.1}^{+30.1}$ & $1424.3_{-27.8}^{+30.0}$  & $1422.6_{-31.7}^{+27.3}$\\
\hline
$H^{-1}\,(\mpc)$   & $3240.1$ & $3157.5_{-176.1}^{+190.6}$ & $3179.3_{-185.6}^{+198.1}$  & $3218.3_{-173.3}^{+200.3}$\\
\hline
$G_b$   & $-$ & $1.13_{-0.08}^{+0.09}$ & $1.10_{-0.06}^{+0.06}$ & $1.15_{-0.08}^{+0.09}$\\
\hline
$G_{\Theta}$ &$0.46$ & $0.39_{-0.09}^{+0.09}$ &$0.46 {\rm \,(fixed)}$ &$0.41_{-0.09}^{+0.09}$\\
\hline
$\sigma_p\,(\mpc)$ &$-$ & $10.6_{-4.6}^{+4.5}$ & $11.7_{-4.1}^{+3.9}$  & $9.3_{-5.6}^{+5.3}$\\
\hline
$\log |f_{R0}|$ & $-$ &  $-4.5_{\rm [prior]}^{+0.96(1.4)}$ &  $-4.5_{\rm [prior]}^{+0.63(0.90)}$ & $-$\\
\hline
$\chi^2_{\rm min}$ & $-$&126.9&127.6&127.7 \\
\hline
\end{tabular}
\end{center}
\caption{The Planck best-fit $\Lambda$CDM predictions and the measured values 
with their 1-$\sigma$ confidence level errors are shown for the parameters 
$D_A,H^{-1},G_b,G_{\Theta},\sigma_p,\log|f_{R0}|$. The Planck $\Lambda$CDM 
data does not predict the phenomenological parameters $G_b$ and $\sigma_p$. 
The last two columns show the measured values in the test cases when 
$G_{\Theta}$ is fixed to the Planck prediction $0.46$, or $f_{R0}=0$, 
respectively. 
} 
\label{tab:measurements}
\end{table*}


In order to check the validity of our constraints, we test the effect of a different initial broadband shape of power spectra by using the WMAP9 best fit model as fiducial instead of Planck 2013. Although there is no significant difference within statistical allowances, the predicted initial broadband shape is slightly different in both cases. 
We would like to test whether this can cause a shift in the an apparent 
$f_{R0}$. 
The black dashed curve in the right panel of Fig.~\ref{fig:measuredfR} represents the likelihood function and $\Delta\chi^2$. The measured $\log |f_{R0}|$ using WMAP9 prior is $-4.6_{\rm [prior]}^{+0.81(1.2)}$, which is consistent with the result using Planck 2013 prior. Thus our results do not appear to be biased by 
the CMB priors. 

Finally, since we measure $f_{R0}$ through its scale dependent effects, 
we consider a non-trivial scale dependent bias on our measurement. We model the  
bias as
\ba
b(k)=b_0\frac{1+A_2k^2}{1+A_1k} \, .
\ea
This simple bias model, when implemented in the formula in Eq.~(\ref{eq:TNS10}), can nicely fit the N-body data of the halo power spectrum and extract correctly the linear growth rate on scales very relevant to the current analysis~\cite{Nishimichi:2011jm,Ishikawa:2013aea}.
We find that the scale dependent bias parameters are determined as 
$A_2 = -0.8{\pm 4}$ and $A_1 = -0.2{\pm 1.2}$. Given the range of errors on these parameters and the range of wavenumbers $k\la 0.1\ompc$, the scale dependence of 
the bias is poorly determined. Since our established cutoff on small scales removes where scale dependent bias is expected to be most significant, all other measurements, including $f_{R0}$, remain similar to the results with scale independent bias. Thus the measurements shown in Table~\ref{tab:measurements} are not modified much by a possible existence of this scale dependent galaxy bias. 

\begin{figure*}
\begin{center}
\resizebox{3.4in}{!}{\includegraphics{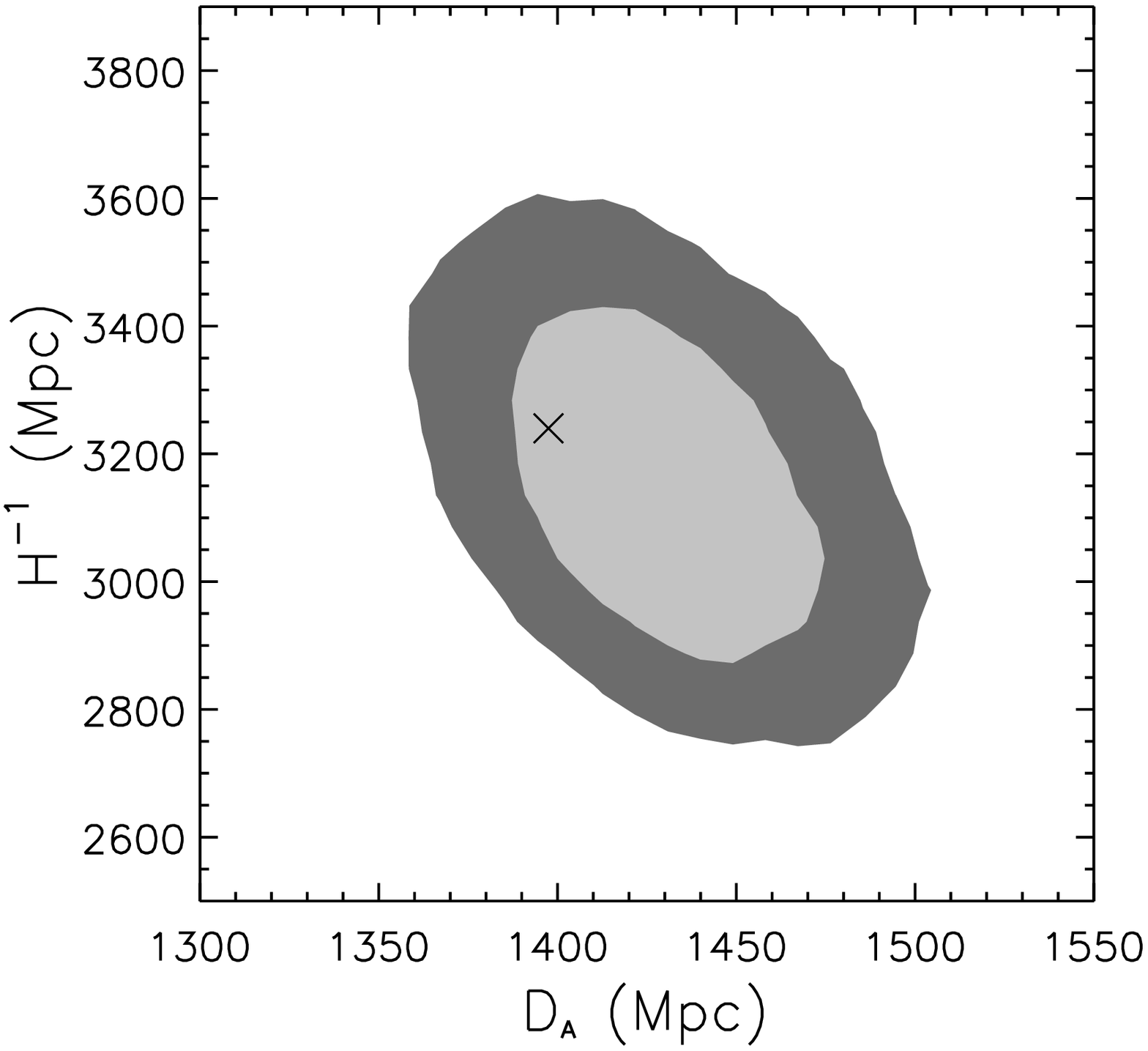}}
\resizebox{3.4in}{!}{\includegraphics{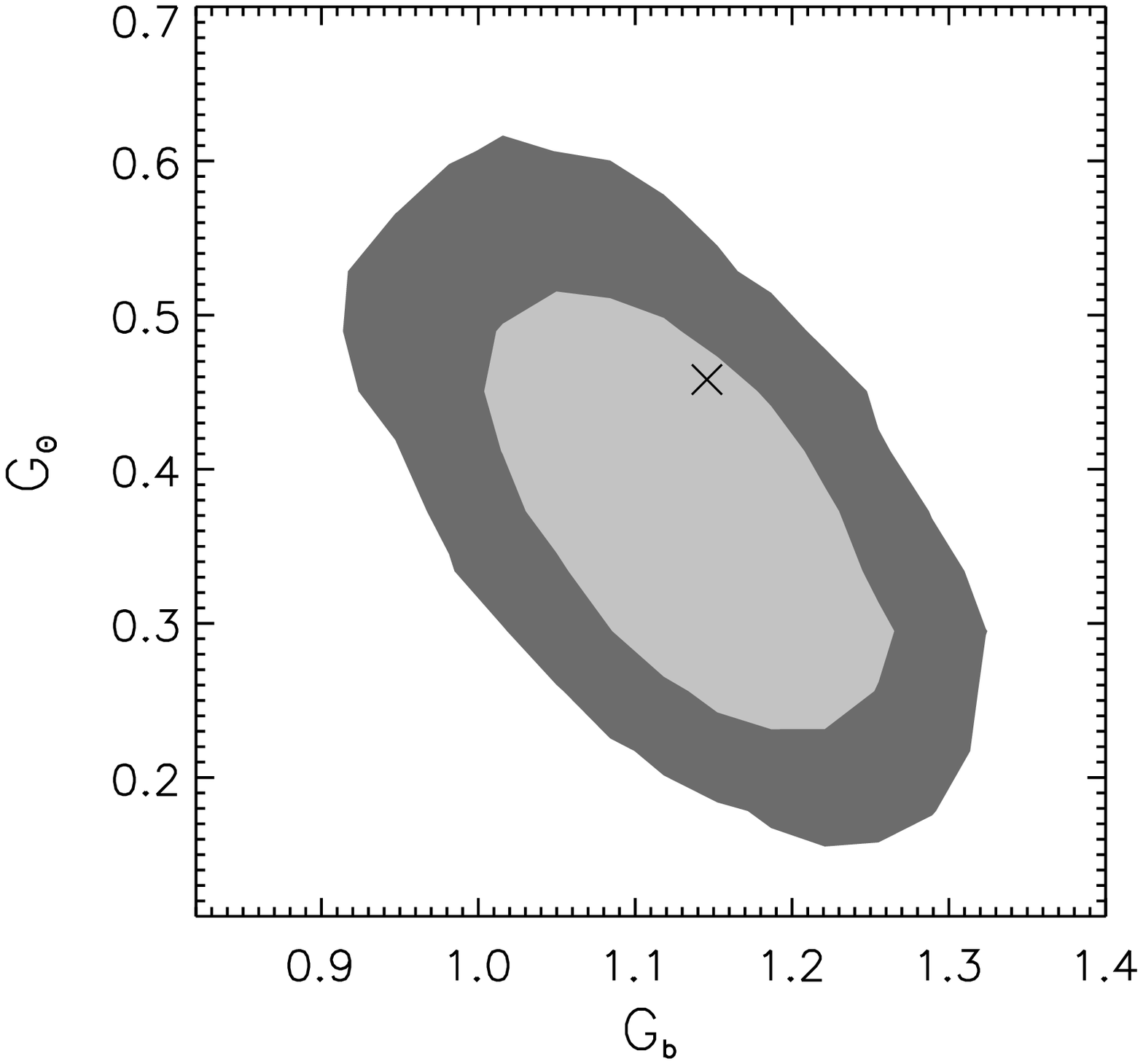}}\\ 
\resizebox{3.4in}{!}{\includegraphics{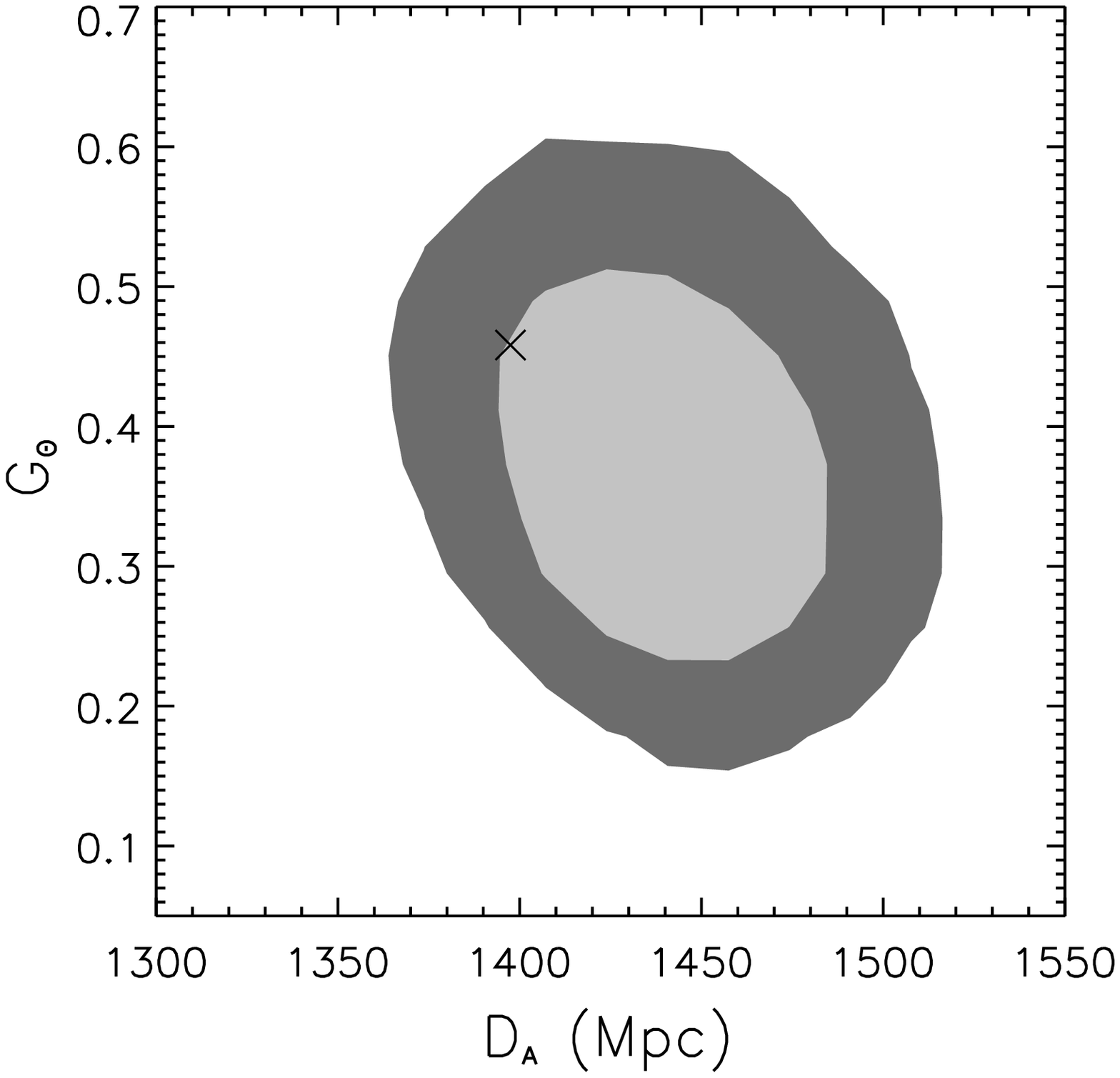}}
\resizebox{3.4in}{!}{\includegraphics{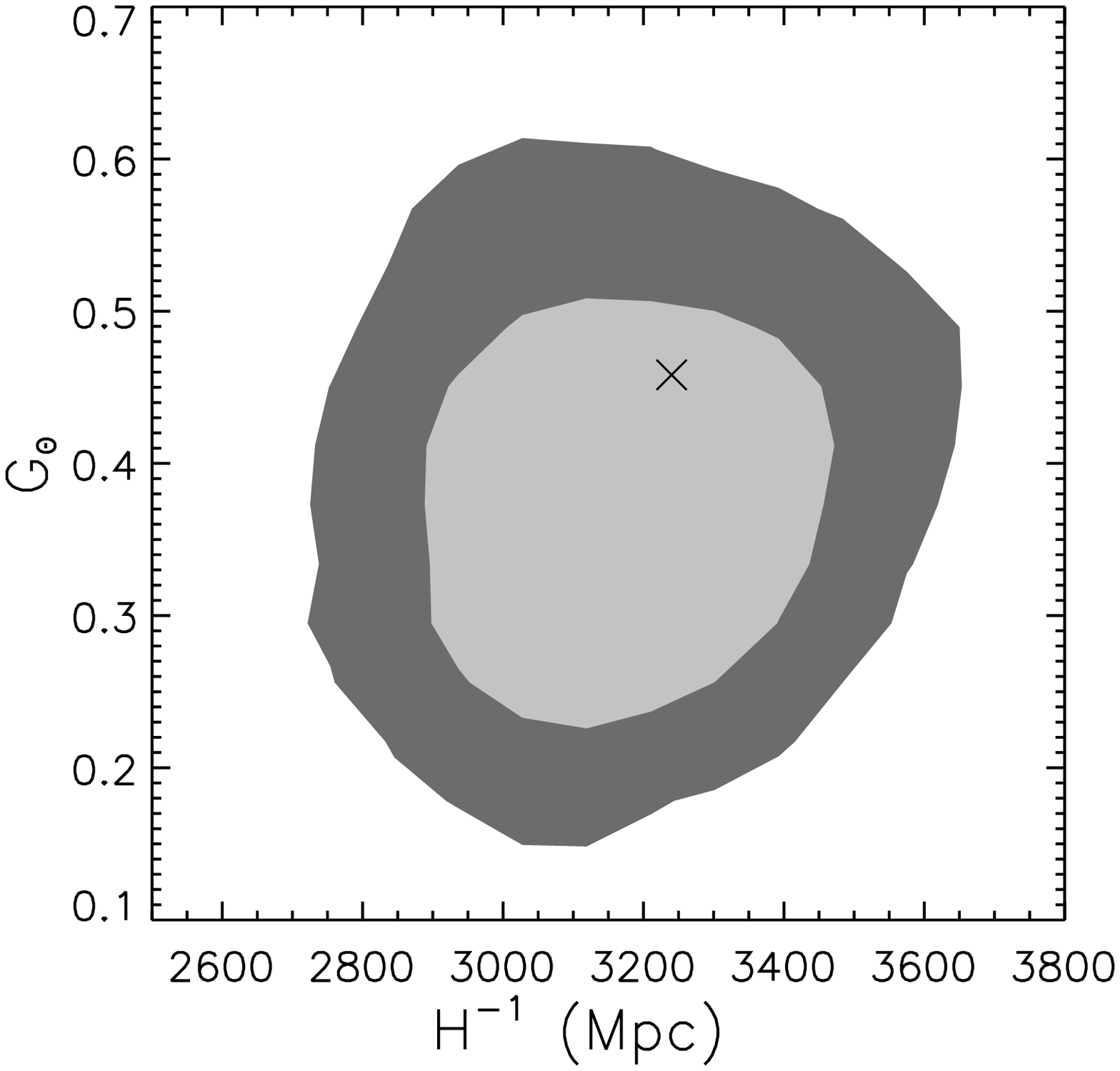}}
\end{center}
\caption{The 2D joint likelihood contours at 68\% and 95\% CL measured for $D_A$, $H^{-1}$, $G_b$ and $G_{\Theta}$ are shown, using  $s_{\rm cut}=50\hompc$ and $\sigma_{\rm cut}=40\hompc$. 
The x denotes the Planck best-fit $\Lambda$CDM predictions (with $G_b$ taking 
the value from fixing $f_{R0}=0$). 
} 
\label{fig:2Dparam}
\end{figure*}

\subsection{Tests of $\Lambda$CDM model}\label{sec:cosmo2} 

Considering the constraints on the other fitting parameters, we can see if there is any evidence of deviations from $\Lambda$CDM model. 
Information on the late-time cosmological expansion is encoded in the distances along and transverse to the line of sight. The distance measurements are not affected by the scale dependent growth functions in $f(R)$ gravity models. In section III, 
we varied the scale dependent growth function $D^+_\delta$ and the growth rate $D^+_\Theta$ and found that the BAO ring is invariant (see  Fig.~\ref{fig:xivariation_test}). As a consequence, there is little change in the measured $D_A$ and $H^{-1}$ from our previous result where scale independent growth functions are assumed~\cite{2014JCAP...12..005S}. The measured $D_A$ is $D_A=1428.0_{-28.1}^{+30.1}\mpc$, and the measured $H^{-1}$ is $H^{-1}=3157.5_{-176.1}^{+190.6}\mpc$. The prediction 
of the $\Lambda$CDM model is indicated by the x in Fig.~\ref{fig:2Dparam}, and the measured values, with uncertainties, are presented as filled contours in the top-left panel. The $\Lambda$CDM prediction is within the 1-$\sigma$ confidence level. The measured distances are consistent with the results using $\Lambda$CDM templates presented in the fifth column of Table~\ref{tab:measurements}. 

Although we vary galaxy bias $b$ in the fitting procedure, it is degenerate with the variation of $G_\delta$ on quasilinear scales. Thus as previously stated we 
can regard the combined measurement of $b$ and $G_\delta$ as a probe of $G_b$, whose measured value is $1.13_{-0.08}^{+0.09}$. Again there is little difference from the previous result of $G_b$ using $\Lambda$CDM templates presented in the fifth column of Table~\ref{tab:measurements}. 

\begin{figure*}
\begin{center}
\resizebox{3.4in}{!}{\includegraphics{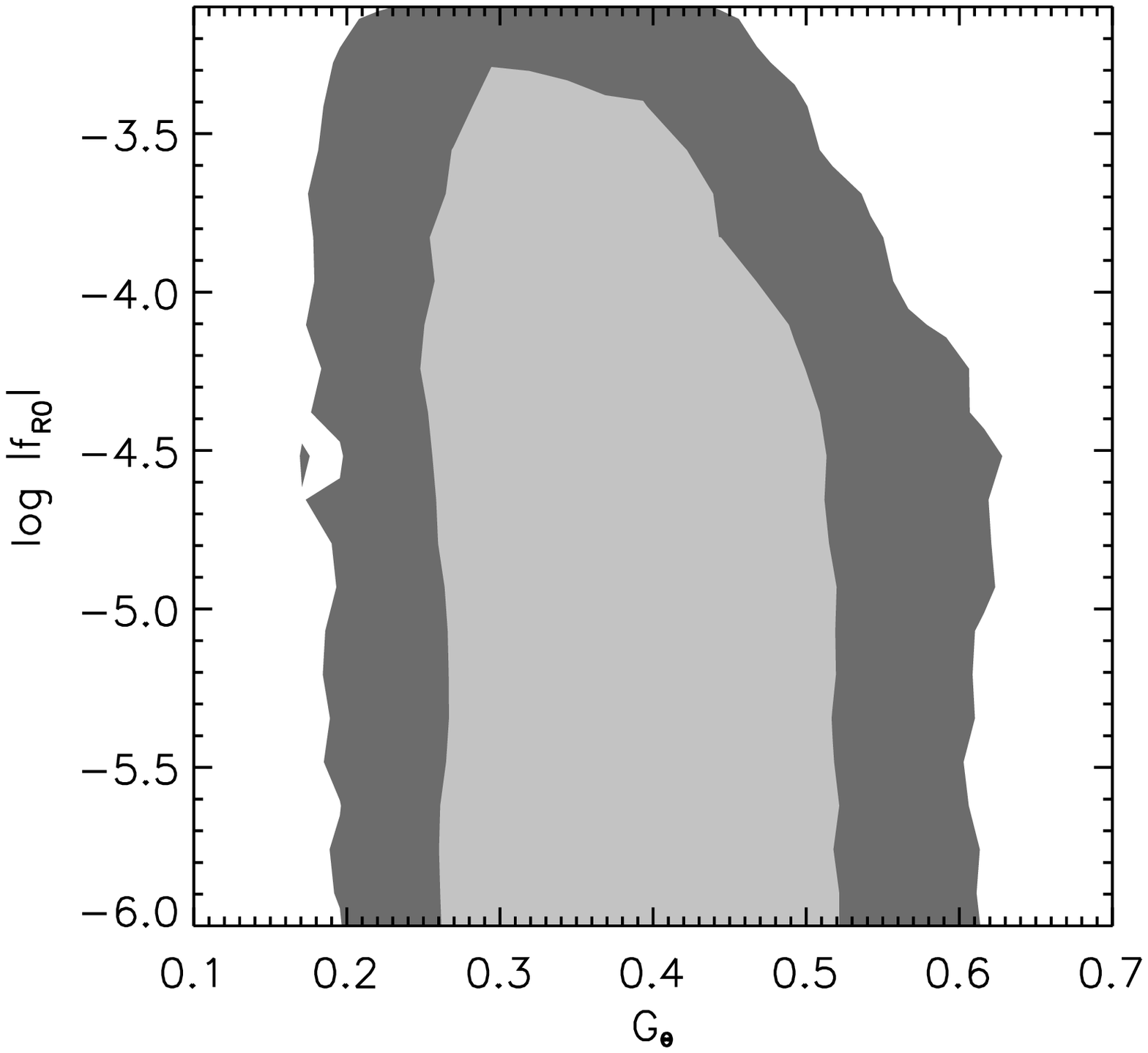}}
\resizebox{3.4in}{!}{\includegraphics{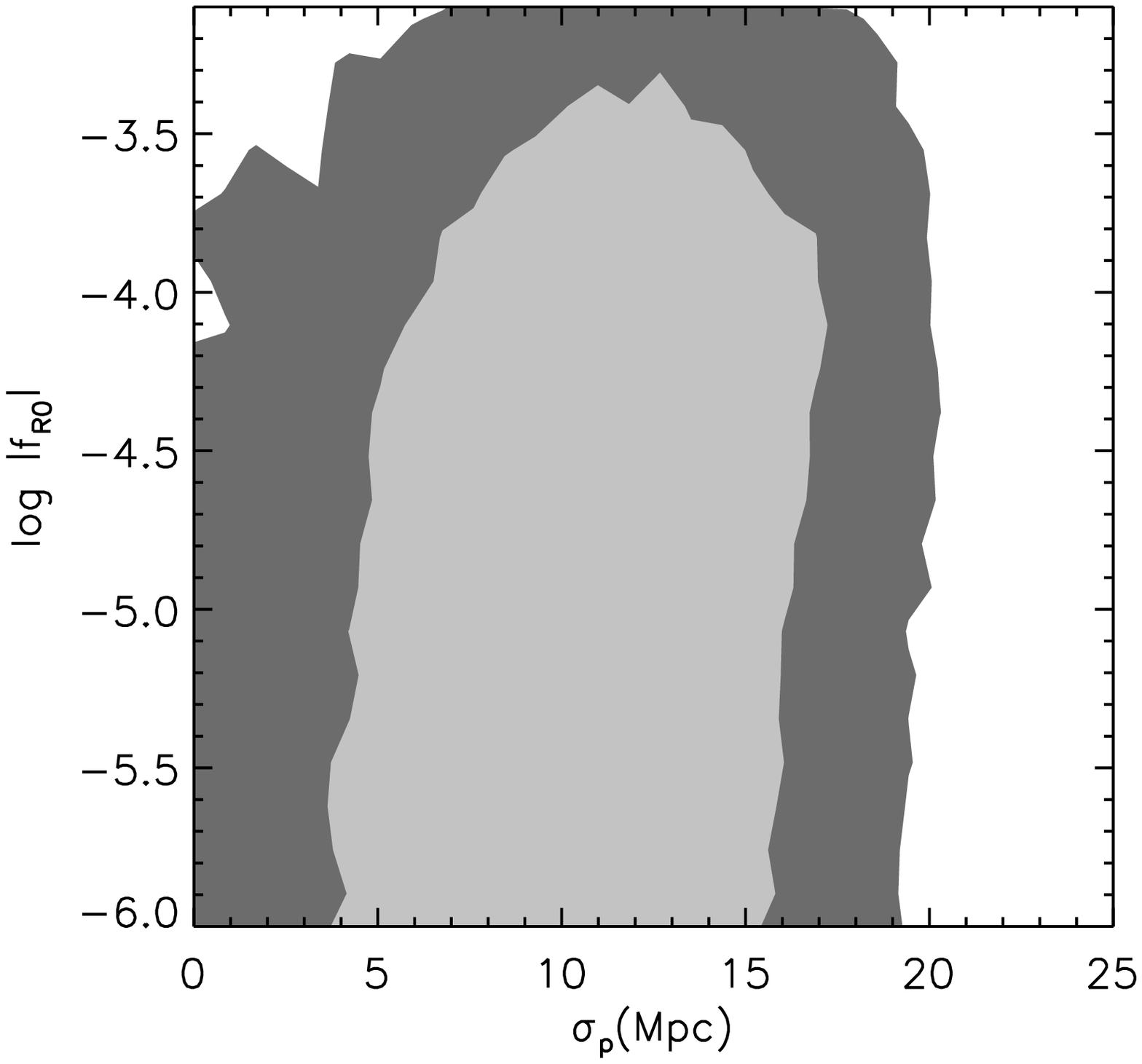}}
\end{center}
\caption{The 2D joint likelihood contours at 68\% and 95\% CL measured for $\log|f_{R0}|$ and $G_{\Theta}$ are shown in the left panel, and for 
$\log|f_{R0}|$ and $\sigma_p$ are shown in the right panel, using  $s_{\rm cut}=50\mpcoh$ and $\sigma_{\rm cut}=40\mpcoh$. 
} 
\label{fig:con_fR_GT}
\end{figure*}

When we fit the measured $\xi(\sigma,\pi)$, the amplitudes of the growth function and growth rate are determined at an effective scale $k^*\sim 0.07\hompc$ in our analysis. In $f(R)$ gravity the growth functions are enhanced in a scale dependent 
manner. Therefore if the amplitude of the growth function and growth rate at $k^*$ are tuned to be nearly the same as the $\Lambda$CDM model (e.g.\ to agree with 
data), then the measured $G_b$ and $G_{\Theta}$ must be smaller to offset the 
enhancement. So for appreciable $\log|f_{R0}|$ we 
expect an anti-correlation between $\log |f_{R0}|$ and $G_{\Theta}$. 
This is visible in the left panel of Fig.~\ref{fig:con_fR_GT} for the larger 
values of $\log |f_{R0}|$. 
Since the best fit $\log |f_{R0}|$ is slightly larger than the LCDM bound, 
this may contribute to why the measured $G_{\Theta}=0.39_{-0.09}^{+0.09}$ 
is smaller than the $\Lambda$CDM result. 
However, the prediction of $\Lambda$CDM presented by the x in 
Fig.~\ref{fig:2Dparam} is still within the 68\% CL measured 
$G_{\Theta}$ contours. 

Finally, we show the measured FoG effect in terms of $\sigma_p$ in the right panel of Fig.~\ref{fig:con_fR_GT}. Again we do not find any significant change from the previous result using $\Lambda$CDM templates. 

\section{Conclusion}\label{sec:concl} 

We have investigated the effect of $f(R)$ gravity on the redshift space correlation function $\xi(\sigma,\pi)$ using a consistent modified gravity perturbation 
theory template that also includes some nonlinear and chameleon screening 
effects. The scale dependent growth functions $D^+_\delta$ and $D^+_\Theta$ alter at late times the initial broadband shape probed by the CMB experiments, but the comoving scale of the sound horizon is not affected, and 
we find the measured BAO ring remains the same as in $\Lambda$CDM and so the 
distances $D_A$ and $H^{-1}$ do also. The effect of modified gravity would 
be most evident in the redshift space distortions and hence anisotropy of the 
clustering. 

We demonstrated how $\xi(\sigma,\pi)$ alters with increments of $D^+_\delta$ and $D^+_\Theta$ for small and large $f_{R0}$ values in Fig.~\ref{fig:xivariation_test}. This shifts the correlation amplitude at the BAO scale and other scales 
compared with the scale independent growth functions case. At small 
$|f_{R0}|=3.2\times 10^{-5}$, there is no significant anisotropic variation 
with $\Delta D^+_\Theta$; 
the peak locations move nearly coherently anti--clockwise along the BAO ring. 
But when $|f_{R0}|$ increases and it becomes larger than $|f_{R0}|= 10^{-4}$, the effect of anisotropic amplification dominates and $\xi(\sigma,\pi)$ departs from $\Lambda$CDM. 

Using the BOSS DR11 data we measure the anisotropic $\xi(\sigma,\pi)$, 
presented as blue filled contours in Fig.~\ref{fig:DR11}, and find a best fit 
$|f_{R0}|=3.2\times 10^{-5}$, with 
an upper bound $|f_{R0}|<3.0\times 10^{-4}$ at the 68\% confidence level; the 
lower 68\% confidence level is consistent with $\Lambda$CDM. 
We also note an anticorrelation between $f_{R0}$ and the scale independent 
growth rate $G_\Theta$ that could reduce the measured value of $G_\Theta$ 
despite the enhancement of growth due to $f(R)$ gravity. 

We tested our data analysis pipeline by applying it to $\Lambda$CDM mock 
catalogues, and we reproduced the $\Lambda$CDM bound of $|f_{R0}|\la 10^{-6}$. We also tested the robustness of our results against the change of the CMB prior. There is a slight difference in the measured early broadband shape of spectra between WMAP9 and Planck 2013, but the measured $|f_{R0}|$ is independent of this difference. In addition, we tested the effect of the scale dependent bias on the measured $|f_{R0}|$. Again our results were shown to be insensitive to a possible scale dependent galaxy bias. 

Although the best fit value of $|f_{R0}|$ is away from $\Lambda$CDM bound, the deviation is insignificant with $\Delta \chi^2\la 1$, and it is indistinguishable from $\Lambda$CDM model. The comparison of the measured minimum $\chi^2_{\rm min}$ 
when marginalising or fixing $\log |f_{R0}|$ to zero, and marginalising or 
fixing the scale independent growth rate to the Planck best fit, 
are presented in last row of Table~\ref{tab:measurements}. 
With future wide, deep spectroscopy experiments such as Dark Energy 
Spectroscopy Instrument (DESI) we will be able to use the redshift space correlation function $\xi(\sigma,\pi)$ to probe $f(R)$ gravity below $|f_{R0}|<10^{-4}$ 
-- and other scale dependent modified gravity -- and test general relativity more stringently.

\section*{Acknowledgments}

Numerical calculations were performed by using a high performance computing 
cluster in the Korea Astronomy and Space Science Institute. EL was 
supported in part by by US DOE grants DE-SC-0007867 and DE-AC02-05CH11231. 
KK is supported by the Science and Technology Facilities Council (grant number K00090X/1). TN was supported by JSPS Postdoctoral Fellowships for Research Abroad. TO was supported by Grant-in-Aid for Young Scientists (Start-up) from the Japan Society for the Promotion of Science (JSPS) (No. 26887012). 
GBZ is supported by the Strategic Priority
Research Program "The Emergence of Cosmological
Structures" of the Chinese Academy of Sciences Grant
No. XDB09000000, and by University of Portsmouth. 
We thank Marc Manera for providing the mock simulations.

Funding for SDSS-III has been provided by the Alfred P. Sloan Foundation, the Participating Institutions, the National Science Foundation, and the U.S. Department of Energy Office of Science. The SDSS-III web site is http://www.sdss3.org/. 

SDSS-III is managed by the Astrophysical Research Consortium for the Participating Institutions of the SDSS-III Collaboration including the University of Arizona, the Brazilian Participation Group, Brookhaven National Laboratory, Carnegie Mellon University, University of Florida, the French Participation Group, the German Participation Group, Harvard University, the Instituto de Astrofisica de Canarias, the Michigan State/Notre Dame/JINA Participation Group, Johns Hopkins University, Lawrence Berkeley National Laboratory, Max Planck Institute for Astrophysics, Max Planck Institute for Extraterrestrial Physics, New Mexico State University, New York University, Ohio State University, Pennsylvania State University, University of Portsmouth, Princeton University, the Spanish Participation Group, University of Tokyo, University of Utah, Vanderbilt University, University of Virginia, University of Washington, and Yale University.

\appendix
\section{Power spectrum and bispectrum calculations in RegPT}
\label{appendix:regpt}
In this appendix, based on the {\tt RegPT} treatment, 
we summarize the expressions for the basic ingredients 
needed to compute the redshift-space power spectrum in $f(R)$ gravity model 
[Eq.~(\ref{eq:TNS10})]. 
The {\tt RegPT} scheme is based on a multipoint propagator expansion, and with this scheme any statistical quantities consisting of density and velocity fields are built up with multipoint propagators, in which nonperturbative properties of gravitational growth  are wholly encapsulated~\citep{Bernardeau:2008fa}. 
Making use of the analytic properties of the propagators, a novel regularized treatment is constructed with the help of the standard PT kernels~\citep{Bernardeau:2011dp}, showing that the proposed scheme can be used to give a percent-level prediction of the power spectrum and correlation function in the weakly nonlinear regime in both real and redshift spaces~\citep{Taruya:2012ut,Taruya:2013my}. 


Let us define a two-component multiplet, $\Psi_a(\bfk;t)\equiv(\delta(\bfk;t),\,\Theta(\bfk;t))$. Then, the power spectra of $\Psi_a$ valid at 
one-loop order are expressed as
\ba
&&P_{ab}(k;t)=\Gamma^{(1)}_a(k;t)\Gamma^{(1)}_b(k;t) P_0(k) 
\nonumber\\
&&
+\quad 2\,\int\frac{d^3\bfq}{(2\pi)^3} \,
\Gamma^{(2)}_a(\bfq,\bfk-\bfq;t)\Gamma^{(2)}_b(\bfq,\bfk-\bfq;t)
\nonumber\\ 
&& \qquad\times\, 
P_0(q)P_0(|\bfk-\bfq|),
\label{eq:pk_RegPT1loop}
\ea
where $P_0$ is the power spectrum of initial density field $\delta_0$. 
The functions $\Gamma_a^{(n)}$ are the multipoint propagators. 
For the one-loop calculation, the relevant expressions for 
the {\it regularized} propagators, which reproduce 
both the resummed behavior at high-$k$ and standard PT results at low-$k$,  
are given by 
\ba
&&\Gamma^{(1)}_a(k;t)=
\Bigl[F_a^{(1)}(k;t)\left\{1+\frac{k^2\sigmad^2}{2}\right\}
\nonumber\\
&&\quad+
3 \int\frac{d^3\bfq}{(2\pi)^3}F_a^{(3)}(\bfk,\bfq,-\bfq;t)\,P_0(q)\Bigr]
e^{-k^2\sigmad^2/2}
\label{eq:G1reg}
\\
&&\Gamma^{(2)}_a(\bfq,\bfk-\bfq;t)=F_a^{(2)}(\bfq,\bfk-\bfq;t)\,
e^{-k^2\sigmad^2/2},
\label{eq:G2reg}
\ea
where the functions $F_a^{(n)}$ are the standard PT kernels, sometimes written as $F_a^{(n)}=(F_n,\,G_n)$ \citep{Bernardeau_review}. Note 
that the leading-order kernel $F_a^{(1)}$ is related to the linear growth 
factor through $F_a^{(1)}=(D_+^{\delta},\,D_+^{\Theta})$. 
The quantity $\sigmad$ is the dispersion of the linear 
displacement field given by 
\be
\sigmad^2=\int\frac{dq}{6\pi^2} P_0(q)\,\left\{\,D_+^{\delta}(q;t)\,\right\}^2.
\ee

The above expressions are used to compute 
the power spectra $P_{\delta\delta}$, $P_{\delta\Theta}$, and $P_{\Theta\Theta}$ 
that explicitly appear in Eq.~(\ref{eq:TNS10}). 
We also need to evaluate the $A$ and $B$ terms, which implicitly 
depend on the power spectrum and bispectrum 
[see Eqs.~(\ref{eq:Aterm}) and (\ref{eq:Bterm})]. 
Since these terms are regarded as next-to-leading
order, the tree-level calculation 
is sufficient for a consistent one-loop calculation of Eq.~(\ref{eq:TNS10}). 
Thus, the power spectrum and bispectrum in the $A$ and $B$ terms are 
evaluated with 
\ba
&&P_{ab,{\rm tree}}(k)=\Gamma^{(1)}_a(k)\Gamma^{(1)}_b(k)\,P_0(k),
\label{eq:pk_tree}\\
&&B_{abc,{\rm tree}}(\bfk_1,\bfk_2,\bfk_3)=2\Gamma^{(2)}_a(\bfk_2,\bfk_3)
\Gamma^{(1)}_b(k_2)\Gamma^{(1)}_c(k_3)
\nonumber\\
&&\qquad\quad\times\,P_0(k_2)P_0(k_3)\,+\,\mbox{(cyc.perm)} 
\label{eq:bk_tree}
\ea
The propagators $\Gamma_a^{(1)}$ and $\Gamma_a^{(2)}$ in the above are also 
evaluated with the tree-level expressions: 
\be
\Gamma_{a,{\rm tree}}^{(n)}(\bfk_1,\cdots,\bfk_n;t)=
F^{(n)}_a(\bfk_1,\cdots,\bfk_n;t)\,e^{-k^2\sigmad^2/2},
\label{eq:Gamma_tree}
\ee
where $k=|\bfk_{1\cdots n}|$.

Note finally that while the power spectrum and bispectrum given 
above are quite general and valid in any model of modified gravity, 
the expressions for the propagators in {\tt RegPT} might receive 
some corrections. 
This issue has been discussed in detail in Ref.~\cite{Taruya:2014faa}. However, it 
has been  found that within the $f(R)$ gravity model, any corrections due
to the modification of gravity 
are very small and can be neglected at the large scales of our interest. 
Hence, all the expressions given above are basically 
the same as those in GR, except for the standard PT kernels, $F_a^{(n)}$.  
Unlike the GR case, the time and scale dependence of $F_a^{(n)}$ are not 
separately treated in $f(R)$ gravity model. The functional form of $F_a^{(n)}$
is thus non-trivial, even if 
the deviation of gravity from GR is small. In this paper, on the basis of 
Eqs.~(\ref{poisson}), (\ref{scalar}), (\ref{continuity}) and (\ref{Euler}), 
we derive the equations that govern the standard PT kernels. 
Solving these equations numerically, we construct the kernels that are 
tabulated as function of wave vectors \citep{Taruya_inprep}.

\end{document}